\documentclass[structabstract]{aa}  
\usepackage{graphicx}
\usepackage{txfonts}
\def\Chan{{\sl Chandra}}
\def\XMM{XMM-{\sl Newton}}
\def\COS{HST-COS}
\def\CII{C\,{\sc ii}}
\def\CIV{C\,{\sc iv}}
\def\CV{C\,{\sc v}}
\def\CVI{C\,{\sc vi}}
\def\NV{N\,{\sc v}}
\def\NVI{N\,{\sc vi}}
\def\NVII{N\,{\sc vii}}

\def\OIII{O\,{\sc iii}}
\def\OIV{O\,{\sc iv}}
\def\OV{O\,{\sc v}}
\def\OVI{O\,{\sc vi}}
\def\OVII{O\,{\sc vii}}
\def\OVIII{O\,{\sc viii}}
\def\NeVIII{Ne\,{\sc viii}}
\def\NeIX{Ne\,{\sc ix}}
\def\NeX{Ne\,{\sc x}}
\def\MgII{Mg\,{\sc ii}}
\def\MgXI{Mg\,{\sc xi}}
\def\MgXII{Mg\,{\sc xii}}
\def\AlXIII{Al\,{\sc xiii}}
\def\SiX{Si\,{\sc x}}
\def\SiXI{Si\,{\sc xi}}
\def\SiXIV{Si\,{\sc xiv}}
\def\SVIII{S\,{\sc viii}}
\def\SIX{S\,{\sc ix}}
\def\SX{S\,{\sc x}}
\def\SXI{S\,{\sc xi}}
\def\SXII{S\,{\sc xii}}
\def\SXIII{S\,{\sc xiii}}
\def\SXIV{S\,{\sc xiv}}
\def\SXV{S\,{\sc xv}}
\def\ArIX{Ar\,{\sc ix}}
\def\ArX{Ar\,{\sc x}}
\def\ArXI{Ar\,{\sc xi}}
\def\ArXII{Ar\,{\sc xii}}

\def\CaXIV{Ca\,{\sc xiv}}
\def\FeI{Fe\,{\sc i}}
\def\FeII{Fe\,{\sc ii}}
\def\FeIII{Fe\,{\sc iii}}
\def\FeIV{Fe\,{\sc iv}}
\def\FeV{Fe\,{\sc v}}
\def\FeVI{Fe\,{\sc vi}}
\def\FeVII{Fe\,{\sc vii}}
\def\FeVIII{Fe\,{\sc viii}}
\def\FeIX{Fe\,{\sc ix}}
\def\FeX{Fe\,{\sc x}}
\def\FeXI{Fe\,{\sc xi}}
\def\FeXII{Fe\,{\sc xii}}
\def\FeXIII{Fe\,{\sc xiii}}
\def\FeXIV{Fe\,{\sc xiv}}
\def\FeXV{Fe\,{\sc xv}}
\def\FeXVI{Fe\,{\sc xvi}}
\def\FeXVII{Fe\,{\sc xvii}}
\def\FeXVIII{Fe\,{\sc xviii}}
\def\FeXIX{Fe\,{\sc xix}}
\def\FeXX{Fe\,{\sc xx}}
\def\FeXXI{Fe\,{\sc xxi}}
\def\FeXXII{Fe\,{\sc xxii}}
\def\FeXXIII{Fe\,{\sc xxiii}}
\def\FeXXIV{Fe\,{\sc xxiv}}
\begin{document}
   \title{Multiwavelength campaign on Mrk 509}

   \subtitle{V. \Chan{}-LETGS observation of the ionized absorber}

   \author{J. Ebrero\inst{1}
          \and
          G. A. Kriss\inst{2,3}
          \and
          J. S. Kaastra\inst{1,4}
          \and
          R. G. Detmers\inst{1,4}
          \and
          K. C. Steenbrugge\inst{5,6}
          \and
          E. Costantini\inst{1}
          \and
          N. Arav\inst{7}
          \and
          S. Bianchi\inst{8}
          \and
          M. Cappi\inst{9}
          \and
          G. Branduardi-Raymont\inst{10}
          \and
          M. Mehdipour\inst{10}
          \and
          P. O. Petrucci\inst{11}
          \and
          C. Pinto\inst{1}
          \and
          G. Ponti\inst{12}
          }

   \institute{SRON Netherlands Institute for Space Research, 
              Sorbonnelaan 2, 3584 CA, Utrecht, The Netherlands\\
              \email{J.Ebrero.Carrero@sron.nl}
              \and
              Space Telescope Science Institute, 3700 San Martin Drive, Baltimore, MD 21218, USA
              \and
              Department of Physics and Astronomy, The Johns Hopkins University, Baltimore, MD 21218, USA
              \and
              Astronomical Institute, University of Utrecht,
              Postbus 80000, 3508 TA, Utrecht, The Netherlands
              \and
              Instituto de Astronom\'ia, Universidad Cat\'olica del Norte,
              Avenida Angamos 0610, Casilla 1280, Antofagasta, Chile
              \and
              Department of Physics, University of Oxford, Keble Road, Oxford OX1 3RH, UK
              \and
              Department of Physics, Virginia Tech, Blacksburg, VA 24061, USA
              \and
              Dipartimento di Fisica, Universit\`a degli Studi Roma Tre, Via della Vasca Navale 84, 00146, Roma, Italy
              \and
              INAF-IASF Bologna, Via Gobetti 101, I-40129 Bologna, Italy
              \and
              Mullard Space Science Laboratory, University College London, Holmbury St. Mary, Dorking, Surrey, RH5 6NT, UK
              \and
              UJF-Grenoble 1 / CNRS-INSU, Institut de Plan\'etologie et d'Astrophysique de Grenoble (IPAG) UMR 5274, Grenoble, F-38041, France
              \and
              School of Physics and Astronomy, University of Southampton, Highfield, Southampton, SO17 1BJ, UK
             }

   \date{Received <date>; accepted <date>}

 
  \abstract
   {We present here the results of a 180~ks \Chan{}-LETGS observation as part of a large multi-wavelength campaign on Mrk 509.}
   {We study the warm absorber in Mrk 509 and use the data from a simultaneous \COS{} observation in order to assess whether the gas responsible for the UV and X-ray absorption are the same.}
   {We analyzed the LETGS X-ray spectrum of Mrk 509 using the {{\tt SPEX}} fitting package.}
   {We detect several absorption features originating in the ionized absorber of the source, along with resolved emission lines and radiative recombination continua. The absorption features belong to ions with, at least, three distinct ionization degrees. The lowest ionized component is slightly redshifted ($\Delta v = 73$~km~s$^{-1}$) and is not in pressure equilibrium with the others, and therefore it is not likely part of the outflow, possibly belonging to the interstellar medium of the host galaxy. The other components are outflowing at velocities of $-196$ and $-455$~km s$^{-1}$, respectively. The source was observed simultaneously with \COS{}, finding 13 UV kinematic components. At least three of them can be kinematically associated with the observed X-ray components. Based on the \COS{} results and a previous {{\it FUSE}} observation, we find evidence that the UV absorbing gas might be co-located with the X-ray absorbing gas and belong to the same structure.}
   {}


   \authorrunning{J. Ebrero et al.}
   \titlerunning{Multiwavelength campaign on Mrk 509. V.}

   \maketitle
%

\section{Introduction}
\label{intro}

Active galactic nuclei (AGN) are among the most luminous persistent extragalactic sources known in the Universe. A significant fraction of their bolometric luminosity is emitted at X-ray wavelengths, and originates in the innermost regions of these sources, where a supermassive black hole resides. Gravitational accretion of matter onto this black hole is believed to be the mechanism that powers these objects, and the intense radiation field thus generated might drive gas outflows from the nucleus. Indeed, more than 50\% of the Seyfert 1 galaxies presents evidence of a photoionized gas, the so-called warm absorber (WA hereafter), in their soft X-ray spectra (e.g. Reynolds \cite{Rey97}, George et al. \cite{Geo98}, Crenshaw et al. \cite{Cre03}). These WA are usually blueshifted with respect to the systemic velocity of the source with velocities of the order of hundreds of km~s$^{-1}$ (Kaastra et al. \cite{Kaa00}, Kaspi et al. \cite{Kas01}, Kaastra et al. \cite{Kaa02}).

Interestingly, sources that present absorption in the UV also show absorption in X-rays (Crenshaw et al. \cite{Cre99}), which may point to a possible relation between the two phenomena. With the advent of modern medium- and high-resolution spectrographs, the ionization state, kinematic properties and column densities of the absorption features in X-rays could be determined with higher accuracy. However, even though the average outflow velocity measured in the X-rays is generally consistent with the velocity shifts measured in the UV, the spectral resolution in X-rays is insufficient to resolve the many individual kinematic components present in the UV. 

Some authors discussed the possibility that at least a fraction of the UV absorption is produced in the same gas responsible for the X-ray absorption (Mathur et al. \cite{Mat94}, \cite{Mat95}, \cite{Mat99}; Crenshaw \& Kraemer \cite{CK99}, Kriss et al. \cite{Kriss00}, Kraemer et al. \cite{Kra02}), but the results were inconclusive and the connection between the two absorbing components seems to be very complex. The most common source of uncertainty came from the lack of correspondence between the column densities of the species detected in both UV and X-rays. However, more recent high-resolution observations indicate that these values can be reconciled if one assumes a velocity-dependent partial covering in the UV (e.g. Arav et al. \cite{Arav02}, \cite{Arav03}).

Simultaneous UV/X-ray observations are therefore the key to draw a global and unbiased picture of the WA in AGN (see e.g. Costantini \cite{Cos10} for a review). High-resolution UV observations covering a wide range of ionization states are essential to determine whether some of the UV absorbers are indeed associated with the X-ray warm absorber. Low-ionization absorbers, such as \MgII{} and \CII{}, are only detected in the UV domain, while high-ionization ions like \OVIII{} or \MgXI{} can only be detected in X-rays. However, there is a range of ionization states that may produce ions detected in both bands, namely \NV{} and, particularly, \OVI{}. Therefore, several simultaneous or quasi-simultaneous X-ray and UV spectroscopic observations have been carried out on many sources known to host a X-ray WA and UV absorbers, e.g. NGC 3783 (Kaspi et al. \cite{Kas02}, Gabel et al. \cite{Gab03}), NGC 5548 (Crenshaw et al. \cite{Cre03b}, Steenbrugge et al. \cite{Ste05}), Mrk 279 (Arav et al. \cite{Arav05}, Gabel et al. \cite{Gab05}, Costantini et al. \cite{Cos07}).

Mrk 509 is a Seyfert 1/QSO hybrid and is one of the best studied local AGN because of its high luminosity (L(1-1000~Ryd)$\simeq 3.2\times 10^{45}$~erg~s$^{-1}$) and proximity. It is particularly suitable for an extensive multi-wavelength campaign because of its brightness and confirmed presence of an intrinsic WA (Pounds et al. \cite{Pou01}, Yaqoob et al. \cite{Yaq03}, Smith et al. \cite{Smi07}, Detmers et al. \cite{Det10}) in X-rays, as well as a variety of UV absorption lines (Crenshaw et al. \cite{Cre95}, \cite{Cre99}; Savage et al. \cite{Sav97}, Kriss et al. \cite{Kriss00}, Kraemer et al. \cite{Kra03}) and slow variability, which makes it particularly suited to reverberation studies.

This paper is devoted to the analysis of the \Chan{}-LETGS observation of Mrk 509, which is part of an unprecedented multi-wavelength campaign of observations carried out at the end of 2009, aiming at determining the location of the warm absorber and the physics of its ionized outflow. The campaign involved five satellites (\XMM{}, \Chan{}, INTEGRAL, HST and Swift), and two ground-based facilities (WHT and PAIRITEL), thus covering more than five orders of magnitude in frequency. An overview of the campaign and its goals can be found in Kaastra et al. (\cite{Kaa11a}), hereafter Paper I, and references therein.

This paper is organised as follows: in Sect.~\ref{xrays} we briefly present the reduction method applied, in Sect.~\ref{analysis} we describe the different models used to fit the spectrum, while in Sect.~\ref{discussion} we discuss our results and compare them with previous X-ray observations as well as with the simultaneous \COS{} observation. Finally, our conclusions are summarized in Sect.~\ref{conclusions}. The quoted errors refer to 68.3\% confidence level ($\Delta \chi^2 = 1$ for one parameter of interest) unless otherwise stated.


\section{X-ray observations and data reduction}
\label{xrays}

Mrk 509 was observed on two consecutive orbits in December 2009 by the Low Energy Transmission Grating Spectrometer (LETGS, Brinkman et al. \cite{Bri00}) coupled to the HRC-S detector on board of \Chan{} for a total exposure time of $\sim$180~ks (see Table~\ref{obslog}). The LETGS observation was made simultaneously with the \COS{} spectrograph observations (Kriss et al. \cite{Kriss11}). 

The LETGS data were reduced using the standard pipeline (CIAO 4.2) until the level 1.5 event files were created. The rest of the procedure, until the creation of the final level 2 event file, followed the same steps as the standard pipeline regarding the wavelength accuracy determination and effective area generation, but was carried out using an independent procedure first described in Kaastra et al. (\cite{Kaa02}).

\begin{figure}
  \centering
  \includegraphics[width=6.5cm,angle=-90]{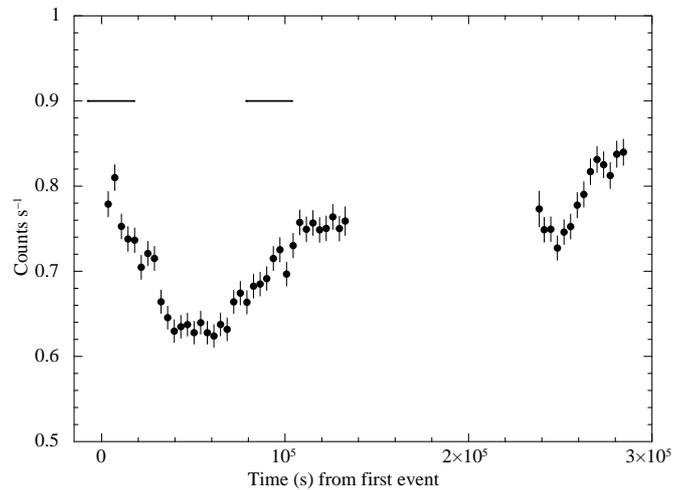}
  \caption{Lightcurve of the \Chan{}~observations. The solid horizontal lines mark the \COS{}~observation intervals.}
  \label{lightcurve}
\end{figure}

The lightcurve of both observations is shown in Fig.~\ref{lightcurve}, with a bin size of one hour (3600~s). We overplot the time intervals when the \COS{} observations were taken. The lightcurve shows no episodes of short-time variability (i.e. within a few kiloseconds) when we apply a bin size of 100~s. At longer timescales the source flux varies smoothly throughout the whole observation. There seems to be a drop in flux, starting shortly after the beginning of the observation, followed by a recovery at $t \sim 50$~ks. The deviation from the average flux was of the order of 30\% and no significant spectral variations were observed between the higher and lower flux states.

\begin{table}
  \centering
  \caption[]{\Chan{}-LETGS observation log.}
  \label{obslog}
  \begin{tabular}{l c c c}
    \hline\hline
    Obs ID  &  Date & Exp. time  & Count rate  \\
            &  (dd-mm-yyyy) &  (ks)  & (counts/s) \\
    \hline
    11387  &  10-12-2009  &  131.4  &  0.94 \\
    11388  &  12-12-2009  &  48.6   &  1.05 \\
    \hline
  \end{tabular}
\end{table}


\section{Spectral analysis}
\label{analysis}

\begin{figure}
  \centering
  \includegraphics[width=6.5cm,angle=-90]{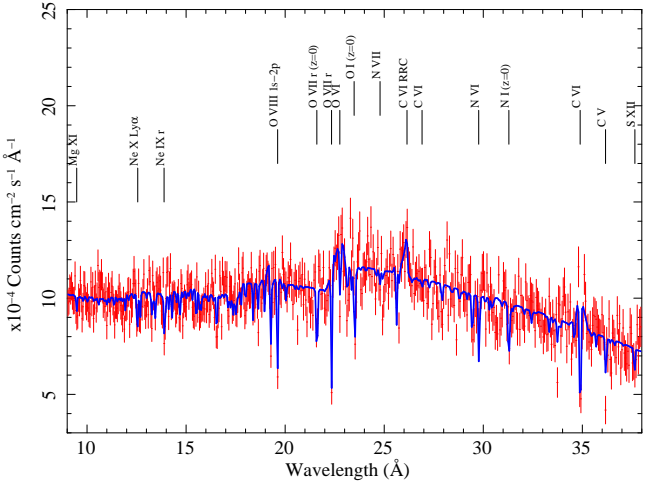}
  \caption{\Chan{}-LETGS spectrum in the $9-38$~\AA~range. The continuum is described with a spline model. The most prominent lines are labeled.}
  \label{fullspec}
\end{figure}

The combined LETGS spectrum was analysed using the {\tt SPEX}\footnote{\tt http://www.sron.nl/spex} fitting package v2.03.00 (Kaastra et al. \cite{Kaa96}). The spectrum has variable bin sizes, depending on the statistics, but is typically binned in 0.03~\AA~bins, which results in $\sim$30~counts per bin. Therefore, the fitting method adopted was $\chi^2$ statistics. In what follows we describe the different models used to describe the continuum, and the absorption and emission features present in the spectrum of Mrk 509. The cosmological redshift of Mrk 509 is $z=0.034397$ (Huchra et al. \cite{Huc93}). For this work, we adopt a systemic redshift of $z=0.03448$, which is the heliocentric redshift of Mrk 509 transferred to the Earth-centered frame of our \Chan{} data at the time of our observation. Note that the standard CIAO software does not take this motion into account. An overall view of the LETGS spectrum of Mrk 509 in the range where the majority of relevant lines fall is shown in Fig.~\ref{fullspec}.

\subsection{Continuum}
\label{cont}

The continuum was modelled following two different approaches. We first used a spline model, that reproduces the shape of the continuum without any {\it a priori} assumptions on its physical origin. The boundaries of the spline were set between 4 and 41~\AA. We note that the analysis of the spectrum is nominally carried out in the range $5-40$~\AA, where the majority of the most relevant spectral features fall. The spline boundaries had to be extended beyond these limits in order to prevent spurious artifacts in the proximity of the boundaries, since the model sets the photon spectrum to zero outside the defined limits. The grid consisted of 13 points spaced logarithmically in wavelength. This spacing is wide enough to successfully model the continuum without risking to fit any intrinsic absorption features. The best-fit flux at the hinge points of the grid is reported in Table~\ref{splinepar}.

As an alternative approach, we also modelled the continuum using a single power-law in the same range as the spline model. This yielded an unacceptable fit, both visually and in terms of reduced $\chi^2$, due to deviations at soft X-rays. To account for these deviations we added a modified black body model ({\tt mbb} model in {\tt SPEX}). The final fit yielded a comparable reduced $\chi^2$ (1742 for 1329 d.o.f.) to that of the spline model (1721 for 1320 d.o.f.). The best-fit parameters are reported in Table~\ref{plmbbpar}. In both cases, the continuum was corrected for local foreground absorption (see Sect.~\ref{local}).

The spline and power-law plus {\tt mbb} unabsorbed models are compared in Fig.~\ref{modelem}, where the rest of emission components are also included. Both models agree well for most part of the fitting range, with average deviations between both models of less than 5\% in the $10-36$~\AA~range. The larger deviations occur below 10~\AA~and above 36~\AA~(where the maximal deviations are only $\sim$7\% and $\sim$9\%, respectively), most likely due to the logarithmic sampling of the spline model. Indeed, the power-law model fails to correctly reproduce the continuum at shorter wavelengths, which is best covered by the spline with 4 hinge points below 10~\AA. The poorer sampling of the spline at longer wavelengths with only one hinge point above 30~\AA~provides a slightly worse description of the continuum than the power-law plus {\tt mbb} model. However, these deviations are lower or similar to the statistical uncertainties in the data, which are of the order of 10\%, and therefore the continuum models do not critically differ. Since the differences between both models in terms of goodness of fit are better for the spline model by $\Delta \chi / \Delta \nu = 21/9$, this was the baseline model adopted for the fit in the $5-40$~\AA~range. The extension of the analysis towards longer wavelengths, up to $\sim$60~\AA, was made using only the power-law baseline model (see Sect.~\ref{WA}).

\begin{table}
  \centering
  \caption[]{Best-fit spline continuum parameters.}
  \label{splinepar}
  \begin{tabular}{l c}
    \hline\hline
    Wavelength (\AA)  &  Flux\tablefootmark{a} \\
    \hline
    4.00  &  $0$ \\
    4.86  &  $11.5 \pm 8.0$ \\
    5.90  &  $11.3 \pm 0.1$ \\
    7.16  &  $10.8 \pm 0.2$ \\
    8.69  &  $11.0 \pm 0.1$ \\
    10.55 &  $11.1 \pm 0.2$ \\
    12.81 &  $12.0 \pm 0.2$ \\
    15.55 &  $12.8 \pm 0.2$ \\
    18.87 &  $14.0 \pm 0.3$ \\
    22.91 &  $15.6 \pm 0.3$ \\
    27.82 &  $17.3 \pm 0.2$ \\
    33.78 &  $19.3 \pm 0.3$ \\
    42.42 &  $0$ \\
    \hline
  \end{tabular}
  \tablefoot{
  \tablefoottext{a}{Unabsorbed flux, in units of $10^{-4}$~photons~cm$^{-2}$~s$^{-1}$~\AA$^{-1}$.}
  }
\end{table}

\begin{table}
  \centering
  \caption[]{Best-fit power-law plus modified black body continuum parameters.}
  \label{plmbbpar}
  \begin{tabular}{l c c c}
    \hline\hline
    Model  &  $\Gamma$  &  $T$\tablefootmark{a}  & $F_{0.5-2}$\tablefootmark{b} \\
    \hline
    Power law & 2.06$\pm$0.03  & $\dots$  & 2.56$\pm$0.05 \\
    Black body & $\dots$  & 0.126$\pm$0.005 & 0.22$\pm$0.03 \\
    \hline
  \end{tabular}
  \tablefoot{
  \tablefoottext{a}{Temperature, in units of keV; \\}
  \tablefoottext{b}{$0.5-2$~keV flux, in units of $10^{-11}$~erg~cm$^{-2}$~s$^{-1}$.}
  }
\end{table}

\subsection{Local absorption}
\label{local}

Before analyzing the spectral features intrinsic to Mrk 509 we first account for the foreground absorption of our Galaxy. The local interstellar medium (ISM) presents three different phases: a cold neutral phase, a warm slightly ionized phase, and a hot highly ionized phase (Pinto et al. \cite{Pin10}). For the neutral phase we take $N_H=4.44 \times 10^{20}$~cm$^{-2}$ (Murphy et al. \cite{Mur96}). The local absorption parameters were kept fixed during the spectral fits. A detailed analysis of the local ISM absorption in the line of sight of Mrk 509 will be reported in a forthcoming paper (Pinto et al., in preparation).

\subsection{Emission lines}
\label{emission}

\begin{figure}
  \centering
  \includegraphics[width=6.5cm,angle=-90]{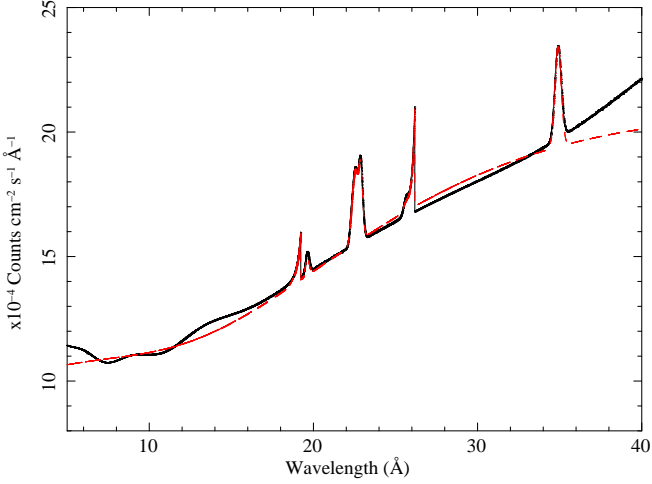}
  \caption{Unabsorbed continuum plus emission line models. Solid line: spline continuum; dot-dashed line: power-law + modified black body.}
  \label{modelem}
\end{figure}

There are several emission features that can be visually identified in the LETGS spectrum of Mrk 509. In most of the cases the emission lines are blended with features from the warm absorber. They need, however, to be modelled in order to account for the excesses seen on both sides of the absorption line. This was the case of  the \CVI{} Ly$\alpha$ line at $\sim$34.9~\AA~and, with less significance, the \OVIII{} Ly$\alpha$ and \NVII{} Ly$\alpha$ lines at $\sim$19.6~\AA~and $\sim$25.5~\AA, respectively. The \OVII{} region shows also great complexity, as a number of emission lines are blended with \OVI{} and \OVII{} absorption lines. Since the lines look  broad, we have modelled the \OVII{} triplet, as well as the lines mentioned above, with Gaussians. The high flux ratio of the \OVII{} intercombination line with respect to the other lines of the \OVII{} triplet suggests that the gas has high density, thus pointing towards an origin in the BLR. This was also found in other Seyfert 1 galaxies (e.g. Longinotti et al. \cite{Lon08}, \cite{Lon09}), and is generally different from the results found in Seyfert 2 galaxies, where the lines are likely produced in the Narrow Line Region (NLR, e.g. Bianchi et al. \cite{Bia06}, Guainazzi \& Bianchi \cite{GB07}). Therefore, the wavelengths were fixed to the rest-frame values and the widths to those of the optical BLR as measured for the Balmer lines with \XMM{} OM ($FWHM \simeq 4000$~km~s$^{-1}$, Mehdipour et al. \cite{Meh11}). The fluxes of the lines are therefore the only free parameters. The \OVII{} forbidden ($f$) and intercombination ($i$), and the \CVI{} lines are detected at the $\sim 3\sigma$ confidence level, whereas the others are just detected at the $1\sigma$ confidence level. The best-fit values are summarised in Table~\ref{gauspar}.

\begin{table}
  \centering
  \caption[]{Best-fit emission line parameters.}
  \label{gauspar}
  \begin{tabular}{l c c c}
    \hline\hline
    Line  &  Wavelength\tablefootmark{a} & FWHM\tablefootmark{b} & Flux\tablefootmark{c} \\
    \hline
    \OVII{} (r)         & 21.602  &  0.285  &  $<$0.43 \\
    \OVII{} (i)         & 21.808  &  0.288  &  0.52$\pm$0.20 \\
    \OVII{} (f)         & 22.101  &  0.292  &  0.67$\pm$0.20 \\
    \NVII{} Ly$\alpha$  & 24.781  &  0.327  &  $<$0.42 \\
    \OVIII{} Ly$\alpha$ & 18.973  &  0.250  &  $<$0.31 \\
    \CVI{} Ly$\alpha$   & 33.736  &  0.445  &  0.67$\pm$0.25 \\
    \hline
  \end{tabular}
  \tablefoot{
  \tablefoottext{a}{Rest wavelength in \AA~(frozen); \\}
  \tablefoottext{b}{Full width at half maximum in \AA~(frozen); \\}
  \tablefoottext{c}{In units of $10^{-4}$~photons~cm$^{-2}$~s$^{-1}$.}
  }
\end{table}

\begin{table}
  \renewcommand{\arraystretch}{1.6}
  \centering
  \caption[]{Best-fit RRC parameters.}
  \label{RRCpar}
  \begin{tabular}{l c c c}
    \hline\hline
    Continuum  & Ion & $EM$\tablefootmark{a} & $T$\tablefootmark{b} \\
    \hline
    Spline      &  \CVI{}  &  7.5$^{+8.9}_{-4.5}$ & 3.7$^{+3.8}_{-2.0}$ \\
                &  \NVII{} &  3.5$^{+4.5}_{-2.1}$ & 7.1$^{+7.4}_{-3.4}$ \\
    \hline
    PL$+$mbb    &  \CVI{}  &  6.2$^{+6.4}_{-3.9}$ & 3.4$^{+3.2}_{-2.1}$ \\
                &  \NVII{} &  4.2$^{+4.2}_{-2.4}$ & 7.9$^{+7.5}_{-3.7}$ \\
    \hline
  \end{tabular}
  \tablefoot{
  \tablefoottext{a}{Emission measure $n_en_{\rm ion}V$, in units of $10^{61}$~cm$^{-3}$; \\}
  \tablefoottext{b}{Temperature, in units of eV.}
  }
\end{table}

In addition to these broad lines, there are also hints of radiative recombination continua (RRC) present in the spectrum. We found a RRC of \CVI{}, seen at $\sim$26.2~\AA, at the $> 2\sigma$ confidence level, and a marginal (detected at the $1.65\sigma$ confidence level) RRC of \NVII{} around 19.1~\AA. Other RRC like that of \OVII{} were statistically undetectable in our dataset. The features were fitted using an ad-hoc model available in {\tt SPEX} that takes into account the characteristic shape of these lines. The best-fit temperature and emission measure of these features are reported in Table~\ref{RRCpar}. The best-fit low temperatures found for these features indicate that the dominant mechanism in the gas is photoionization. The emission line models used in the fit are shown in Fig.~\ref{modelem} for both the spline and power-law continuum models. It can be seen that the choice of either continuum modelling does not significantly affect the detection of the emission lines.

\begin{table*}[!ht]
  \centering
  \caption[]{Ionic column densities of the most relevant ions.}
  \label{slabpar}
  \begin{tabular}{l | c c | c c c | c}
    \hline\hline
             & \multicolumn{2}{c}{Model 1}  & \multicolumn{3}{c}{Model 2} \\  
    \hline
    Component  &  1 &  2 &  1   &  2  &  3  & \\
    $\sigma$\tablefootmark{a} & 108$\pm$11 & 100$^{+63}_{-35}$ & 85$\pm$17  & 61$\pm$15 &  90$\pm$36   &  \\
    $v_{\rm out}$\tablefootmark{b} & 3$^{+17}_{-43}$ & $-338^{+91}_{-142}$ & 87$\pm$36  & $-208\pm$42 & $-635\pm$172    &   \\
    \hline
    Ion  & $\log N_{\rm ion}$\tablefootmark{c} & $\log N_{\rm ion}$\tablefootmark{c} & $\log N_{\rm ion}$\tablefootmark{c} & $\log N_{\rm ion}$\tablefootmark{c} & $\log N_{\rm ion}$\tablefootmark{c}  & $\log \xi$\tablefootmark{d} \\
    \hline
    \CV{}   &  16.47$\pm$0.17 &    & 16.38$\pm$0.19 & 15.95$\pm$0.43 &    & 0.15 \\
    \CVI{}  &  16.66$\pm$0.19 &    & 16.65$\pm$0.18 & 15.85$\pm$0.41 &    & 1.20 \\
    \NV{}   &  $<$15.4        &    &  $<$15.3       & $<$15.4        &    & 0.00 \\
    \NVI{}  & 15.57$\pm$0.36  &    & 15.60$\pm$0.37 & $<$15.3        &    & 0.75 \\
    \NVII{} & 15.89$\pm$0.27  &    & 15.99$\pm$0.23 & $<$15.5        &    & 1.60 \\
    \OIII{} & 16.16$\pm$0.23  &    & 15.96$\pm$0.36 & $<$15.6        &    & $-$1.50 \\
    \OIV{}  & 16.07$\pm$0.31  &    & 15.76$\pm$0.31 & $<$15.9        &    & $-$0.65 \\
    \OV{}   & 16.17$\pm$0.22  &    & $<$15.6        & 16.45$\pm$0.22 &    & $-$0.05 \\
    \OVI{}  & 16.19$\pm$0.18  &    & 15.99$\pm$0.22 & 15.69$\pm$0.39 &    & 0.40 \\
    \OVII{} & 17.10$\pm$0.11  &    & 16.76$\pm$0.14 & 16.54$\pm$0.22 &    & 1.15 \\
    \OVIII{} & 17.43$\pm$0.13 &    & 16.76$\pm$0.31 & 17.14$\pm$0.18 &   & 1.90 \\
    \NeVIII{} & $<$16.0  &    &  $<$16.0       & $<$15.7       &    & 1.20 \\
    \NeIX{} & 16.49$\pm$0.29  &    & 16.54$\pm$0.31 &  $<$15.7       &    & 1.75 \\
    \NeX{}  &    &  16.64$\pm$0.27  & $<$15.8        & 16.37$\pm$0.35 & 16.16$\pm$0.40   & 2.40 \\
    \MgXI{} &    &  16.77$\pm$0.42  & $<$16.1        & $<$16.2        & 16.90$\pm$0.33   & 2.25 \\
    \SiX{} & 15.62$\pm$0.33 &     &   15.40$\pm$0.34 & 15.39$\pm$0.38 &    & 1.90 \\ 
    \SiXI{} &     &  $<$15.2   &  $<$15.2  &  $<$15.1  &     & 2.10 \\
    \SVIII{} & $<$15.2  &     &   $<$15.1   &  $<$15.3   &    & 0.80 \\
    \SIX{} &  15.10$\pm$0.30  &     &  $<$15.2  &  15.17$\pm$0.40 &    & 1.35 \\
    \SX{} &  $<$15.0   &      &  14.75$\pm$0.30  &  $<$14.7  &   & 1.70 \\
    \SXI{} & $<$15.3   &      &  $<$15.5    &  $<$14.8  &    &  1.95 \\
    \SXII{} &    &  $<$15.6   & $<$15.3        & $<$15.7        &    & 2.20 \\
    \SXIII{} &    &  $<$14.9  &  $<$15.4       & $<$14.8        &    & 2.40 \\
    \SXIV{} &    &  $<$15.7   &  $<$15.8       & $<$15.5        &    & 2.60 \\
    \ArIX{} & $<$15.1 &     &   $<$15.0        & $<$15.3        &    & 0.90 \\
    \ArX{}  & $<$15.4 &     &   $<$15.3        & $<$15.2        &    & 1.35 \\
    \ArXI{} & $<$14.4 &     &   $<$14.7        & $<$14.8        &    & 1.70 \\
    \ArXII{} & $<$14.5 &    &   $<$14.5       & $<$14.3        &    & 2.05 \\
    \CaXIV{} &    &  15.53$\pm$0.26   &$<$15.2       & 15.57$\pm$0.24 &    & 2.35 \\
    \FeI{}  & $<$16.0  &    &  $<$15.1        & $<$15.4        &    & $-$8.50 \\
    \FeII{}  & $<$16.1 &    &  $<$15.6       & $<$15.6        &    & $-$8.50 \\
    \FeIII{} & $<$16.1 &    &  15.69$\pm$0.43 & $<$15.5       &    & $-$2.10 \\
    \FeIV{}  & $<$16.0 &    &  $<$15.5       & $<$15.8        &    & $-$1.40 \\
    \FeV{}   & $<$15.3 &    &  $<$15.0       & $<$15.0        &    & $-$0.95 \\
    \FeVI{}  & $<$15.2 &    &  $<$15.2       & $<$15.0        &    & $-$0.55 \\
    \FeVII{} & $<$15.3 &    &  $<$15.0       & $<$15.0        &    & 0.05 \\
    \FeVIII{} & $<$15.6 &    &  $<$15.4      & $<$15.0        &    & 0.85 \\
    \FeIX{}  & 15.44$\pm$0.27 &   &  15.41$\pm$0.26 & $<$15.0       &    & 1.40 \\
    \FeX{}  & $<$15.0  &    &  $<$14.7        & $<$14.9        &    & 1.65 \\
    \FeXI{} & $<$15.5  &    &  $<$15.3        & $<$15.1        &    & 1.85 \\
    \FeXII{} & $<$15.2  &    &  $<$15.2       & $<$15.4        &    & 2.00 \\
    \FeXIII{} & $<$15.3  &    &  $<$15.2      & $<$15.2        &    & 2.05 \\
    \FeXIV{}  &     &  $<$15.5   &$<$15.5      & $<$15.1        &    & 2.10 \\
    \FeXV{}  &      &  $<$15.3    &$<$15.0       & $<$15.0        &    & 2.15 \\
    \FeXVI{} &     &  15.64$\pm$0.22   &$<$15.0       & 15.67$\pm$0.23 &    & 2.20 \\
    \FeXVII{}  &    &  $<$15.3   &$<$14.7     & $<$15.0        &    & 2.30 \\
    \FeXVIII{} &    &  15.84$\pm$0.17  &$<$15.0     & 15.76$\pm$0.19 &    & 2.50 \\
    \FeXIX{}  &    &  $<$15.7   &$<$15.1      & $<$15.6        &    & 2.80 \\
    \FeXX{}  &     &  15.68$\pm$0.31   &$<$15.1       & 15.56$\pm$0.43 &    & 3.00 \\
    \FeXXI{} &    &  15.80$\pm$0.32   &$<$15.6       & $<$15.5        & 15.81$\pm$0.32   & 3.20 \\
    \FeXXII{}  &    &  $<$15.9   & 15.92$\pm$0.50  & $<$15.3    &    & 3.30 \\
    \FeXXIII{} &    &  $<$16.2   &$<$16.2     & $<$15.8        &    & 3.40 \\
    \FeXXIV{} &    &  $<$16.5   & $<$16.5      & $<$16.3        &    & 3.50 \\
    \hline
  \end{tabular}
  \tablefoot{
  \tablefoottext{a}{{\it rms} velocity broadening, in units of km~s$^{-1}$; \\}
  \tablefoottext{b}{Outflow velocity, in units of km~s$^{-1}$; \\}
  \tablefoottext{c}{Ionic column density, in units of cm$^{-2}$; \\}
  \tablefoottext{d}{Ionization parameter at which the concentration of the ions peaks, in units of erg~s$^{-1}$~cm.}
  }
\end{table*}

\subsection{Warm absorber}
\label{WA}

\begin{figure}
  \centering
  \includegraphics[width=6.5cm,angle=-90]{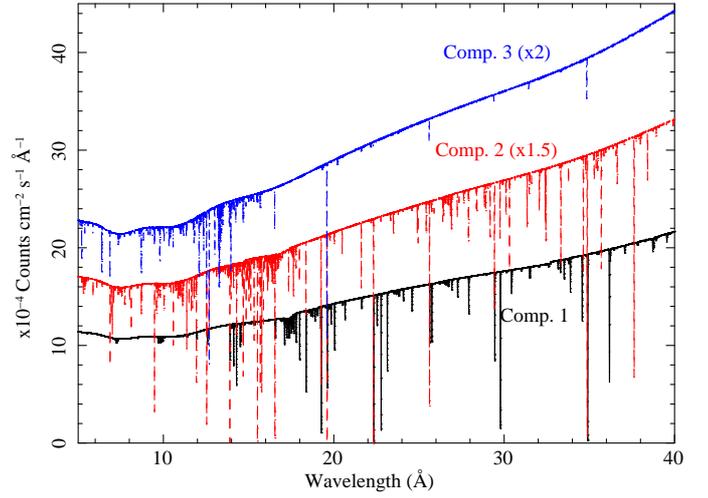}
  \caption{Model plot showing the contribution to the spectrum of each of the three WA components. The solid line represents component 1, while the dashed and dot-dashed lines represent components 2, and 3, respectively . For clarity, components 2 and 3 have been shifted in flux by a factor of 1.5 and 2, respectively.}
  \label{modelxabs}
\end{figure}

The majority of the absorption lines seen in the soft X-ray spectrum of Mrk 509 corresponds to the presence of ionized gas at the redshift of the source. Most of them are blueshifted with respect to the systemic velocity of Mrk 509.

We first modelled the WA using two {\tt slab} components to describe the low- and high-ionization absorption lines, respectively. The {\tt slab} model in {\tt SPEX} calculates the transmission of a slab of material without any assumptions on the underlying ionization balance. The free parameters are the ionic column densities, the {\it rms} velocity broadening, and the outflow velocity. The {\tt slab} fit showed that the low-ionization ions, i.e. those for which their concentrations peak at $\log \xi$ lower than 2.1, appear to be in a different kinematic regime with respect to the high-ionization ions. The ionization parameter is defined as $\xi=L/nR^2$, where $L$ is the ionizing luminosity in the $1-1000$~Ryd range, $n$ is the density of the gas, and $R$ is the distance from the ionizing source. Ions belonging to the first {\tt slab} component are at the systemic velocity, whereas those of component 2 are outflowing at several hundreds of km~s$^{-1}$ (model 1 in Table~\ref{slabpar}). This model is, however, not entirely realistic, as it does not allow for a possible multi-velocity structure in the outflow. Different kinematic components may, in principle, contribute to the total column density of a given ion. Therefore, the model was extended and all the ions were fitted with both velocity components (model 2 in Table~\ref{slabpar}). The results show that indeed \CV{} and \CVI{}, as well as \OVI{}, \OVII{} and \OVIII{} ions are significantly detected in both velocity regimes. Moreover, a third component is required in order to correctly fit \MgXI{} and \FeXXI{} which seem to be outflowing at even larger velocities. Hence, the final fit shows three distinct kinematic components: one slightly redshifted and two blueshifed with respect to the systemic velocity of the source by $\sim$200 and $\sim$600~km~s$^{-1}$. The best-fit values for models 1 and 2 are reported in Table~\ref{slabpar}, along with the ionization parameter $\xi$ for which the ion fraction peaks for that ion. The values of $\xi$ are derived from an ionization balance model and are shown for illustrative purposes, as the {\tt slab} fit is not model-dependent. These different velocity components of the WA will be used as a starting point for a more physical analysis.

We also use an alternative model to describe the WA in Mrk 509. The {\tt xabs} model in {\tt SPEX} calculates the transmission of a slab of material where all ionic column densities are linked through a photoionization balance model, which we calculated using {\it CLOUDY} (Ferland et al. \cite{Fer98}) version C08. The spectral energy distribution (SED) provided to {\it CLOUDY} as an input was built using the different observations involved in this campaign (\Chan{}-LETGS, \COS{}, INTEGRAL, SWIFT), plus archival FUSE data. Details on how the SED was built and plots are shown in Paper I. The free parameters in the {\tt xabs} fit are the ionization parameter $\xi$, the hydrogen column density $N_H$, the line velocity width $\sigma$, and the outflow velocity $v_{out}$. The abundances were set to the proto-Solar values of Lodders \& Palme (\cite{Lod09}).

A single {\tt xabs} component yields a $\chi^2$ of 1490 for 1310 d.o.f. but it is insufficient to describe the different velocity components and the range of ionization states found in the {\tt slab} fit. A second {\tt xabs} component was added, further improving the fit in $\Delta \chi^2/\Delta \nu=46/4$. These two components with $\log \xi \sim 1$ and $\log \xi \sim 2$, respectively, account for the lowly and the majority of the highly ionized ions. However, a few high-ionization lines located at shorter wavelengths (namely \NeIX{}, \FeXX{}, and \FeXXI{}), although weaker, remain unfitted. We therefore added a third {\tt xabs} to our model to fit the fast ($v_{\rm out}\sim -460$~km~s$^{-1}$) high-ionization ($\log \xi \sim 3$) component in the WA. This component is detected slightly above the $2\sigma$ level (the fit improves a further $\Delta \chi^2/\Delta \nu=18/4$), giving also a visually acceptable fit of these ions. Adding more {\tt xabs} components no longer improves the fit, thus we recover three kinematic components in the WA in agreement with the results of the {\tt slab} fit. A plot with the contribution to the spectrum by each {\tt xabs} component is shown in Fig.~\ref{modelxabs}.

\begin{figure*}
  \centering
  \hbox{
    \includegraphics[height=8.5cm,width=5.25cm,angle=-90]{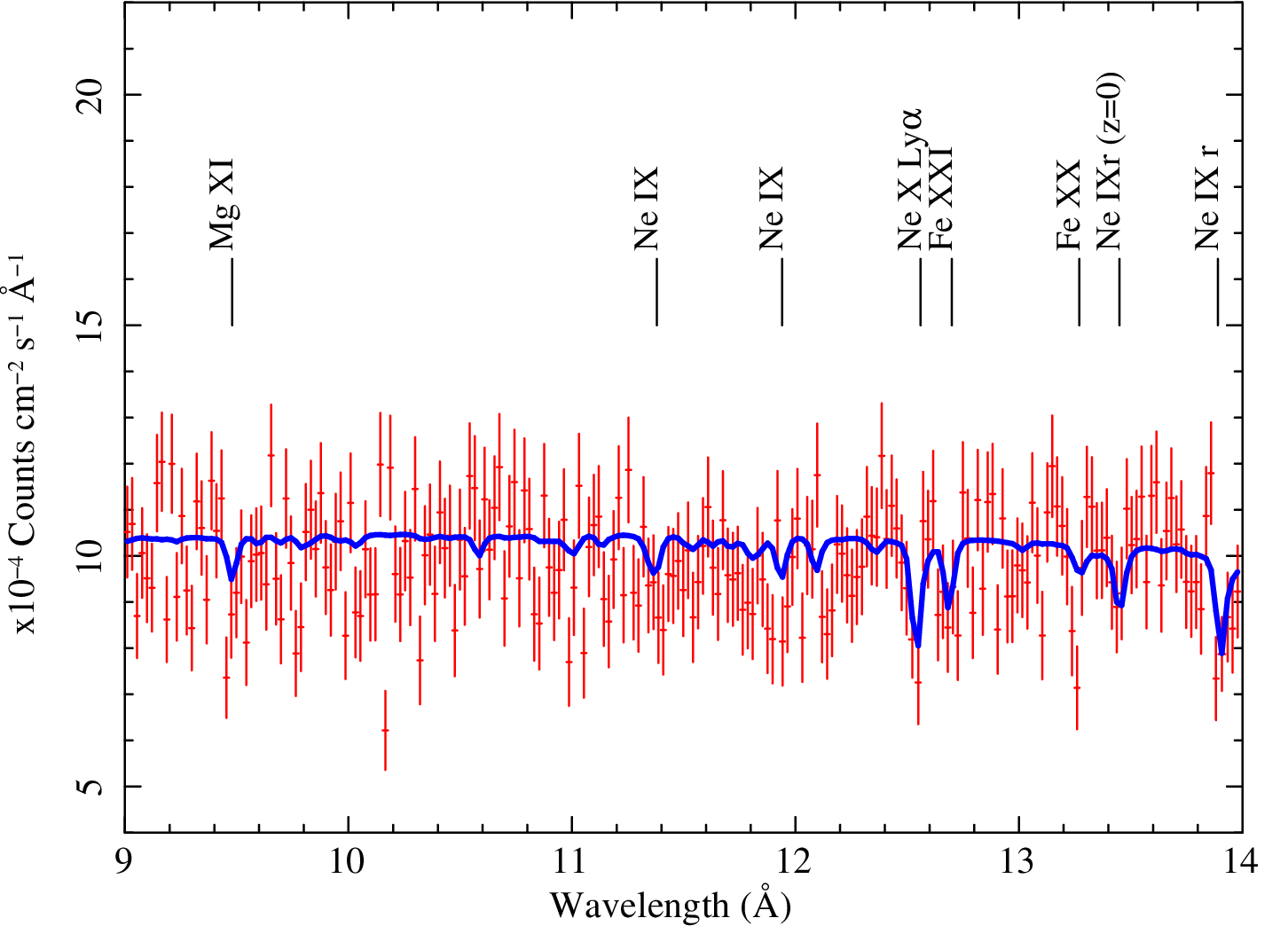}
    \includegraphics[height=8.5cm,width=5.25cm,angle=-90]{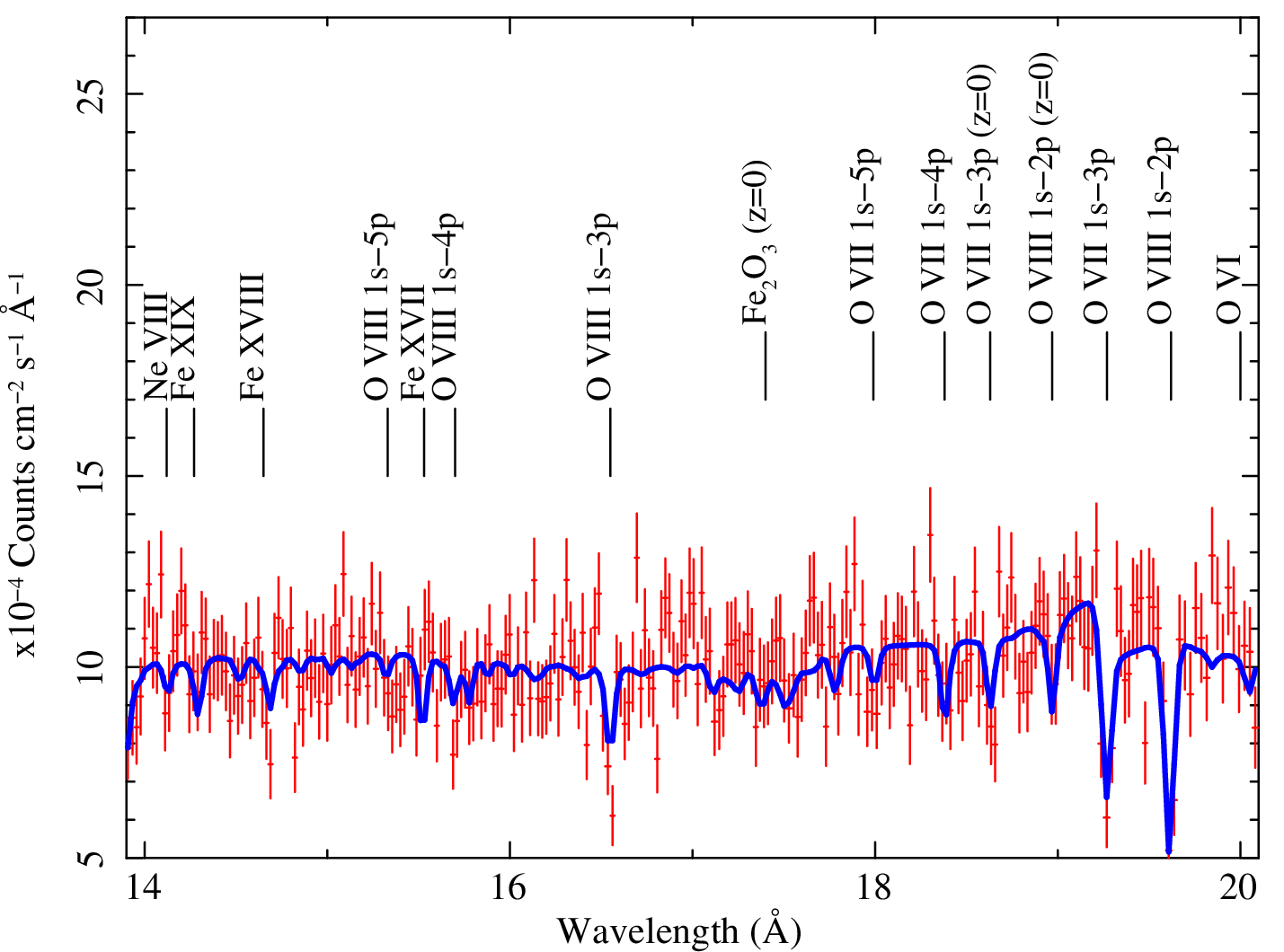}
  }
  \hbox{
    \includegraphics[height=8.5cm,width=5.25cm,angle=-90]{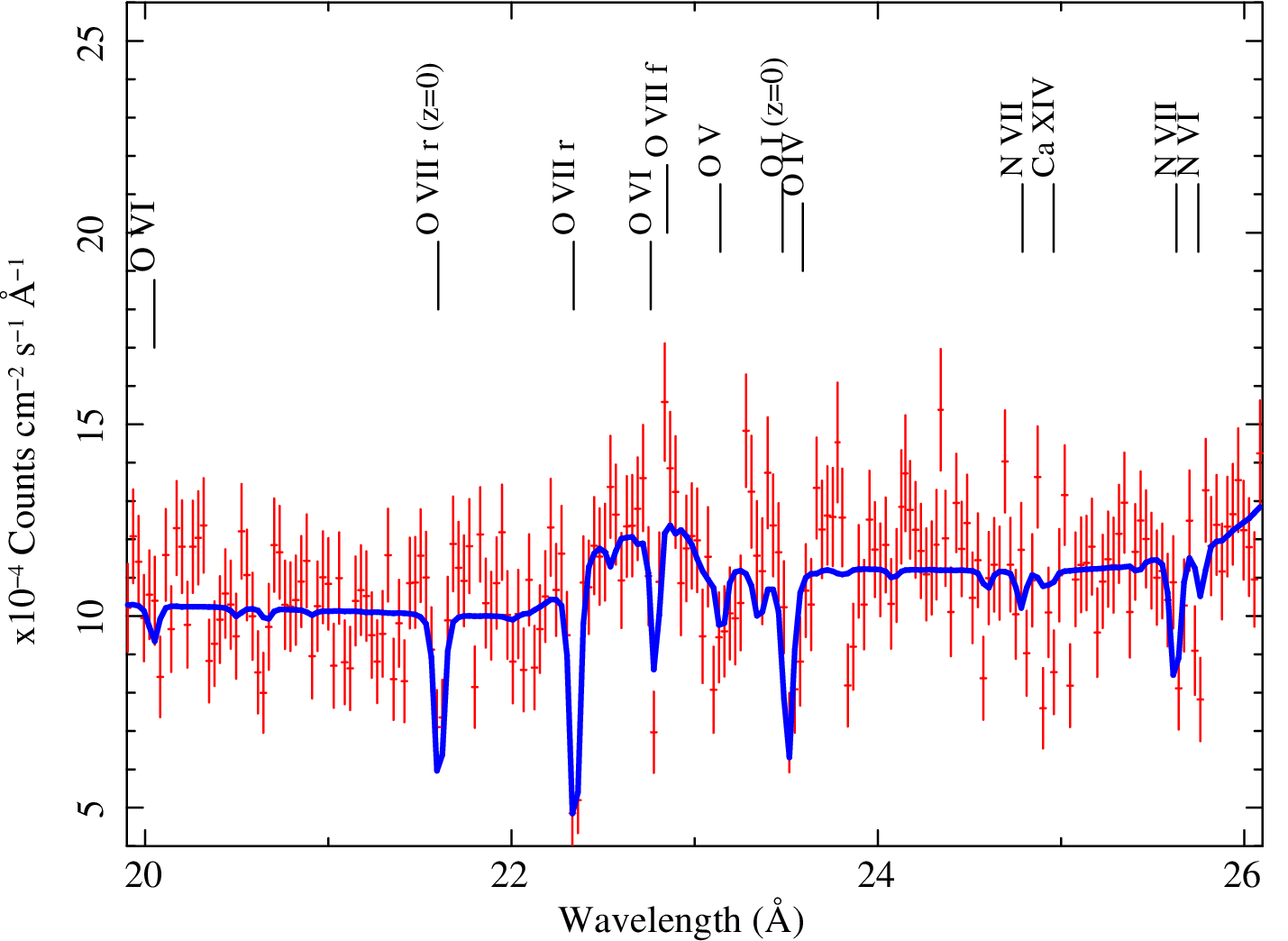}
    \includegraphics[height=8.5cm,width=5.25cm,angle=-90]{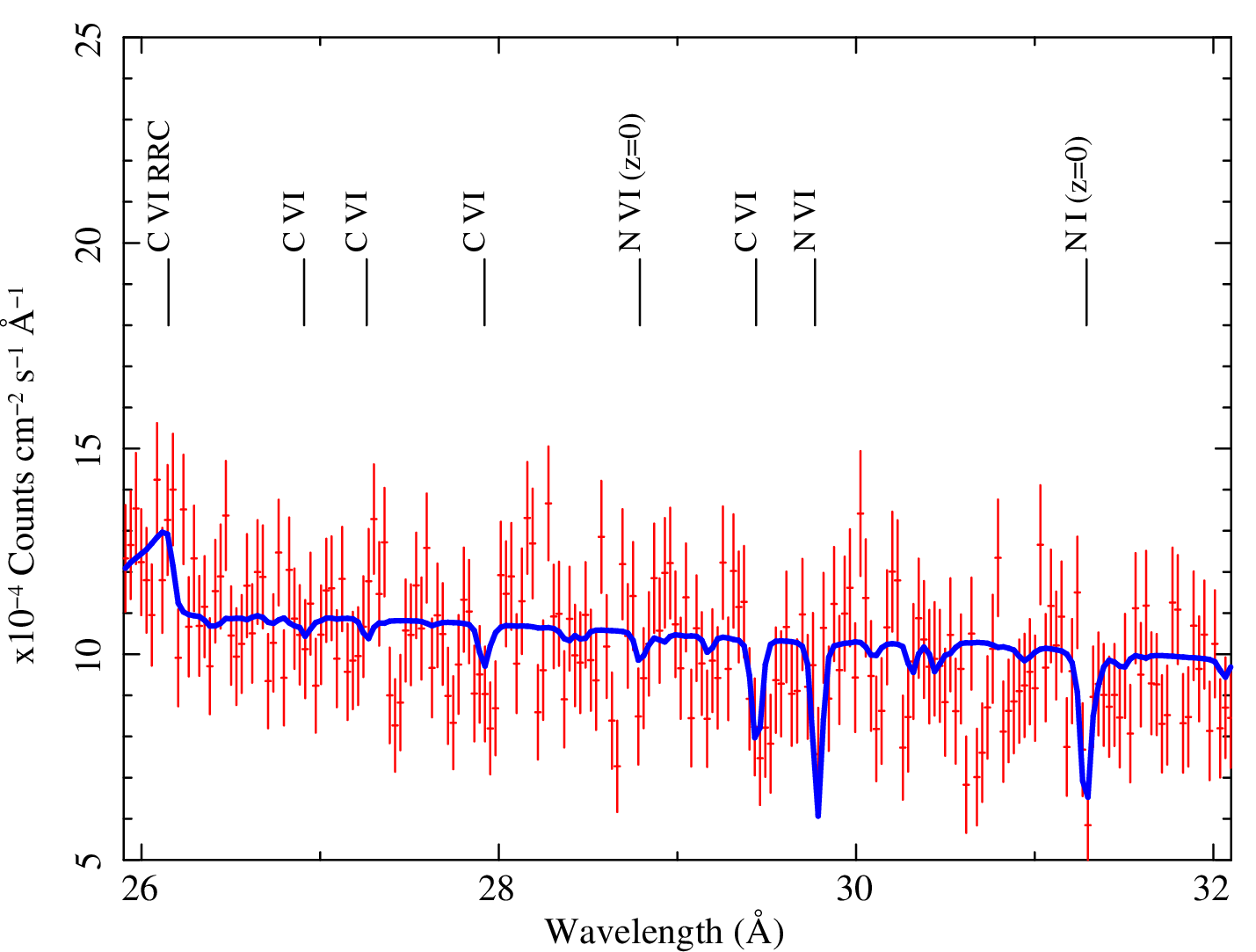}
  }
  \hbox{
    \includegraphics[height=8.5cm,width=5.25cm,angle=-90]{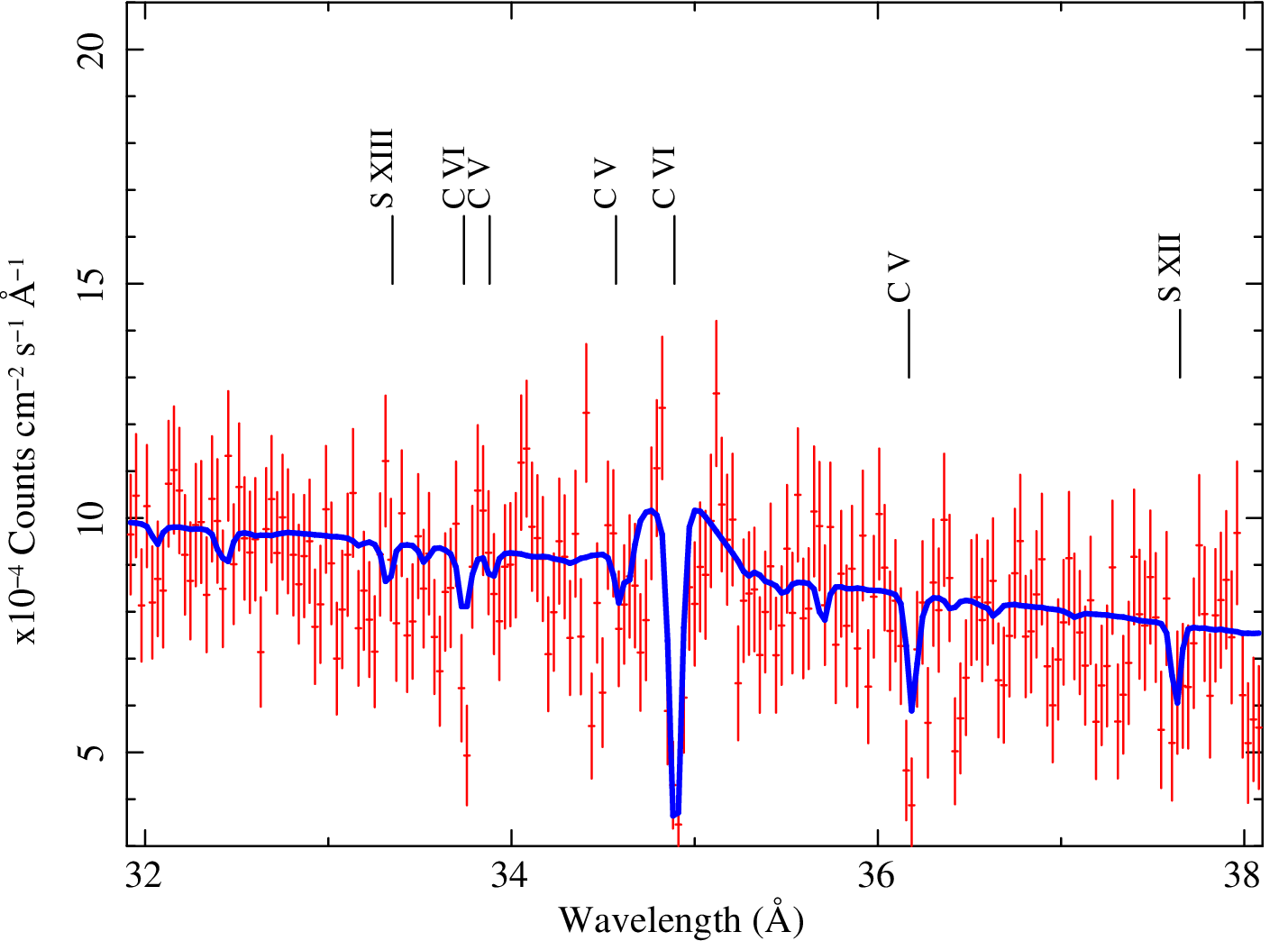}
    \includegraphics[height=8.5cm,width=5.25cm,angle=-90]{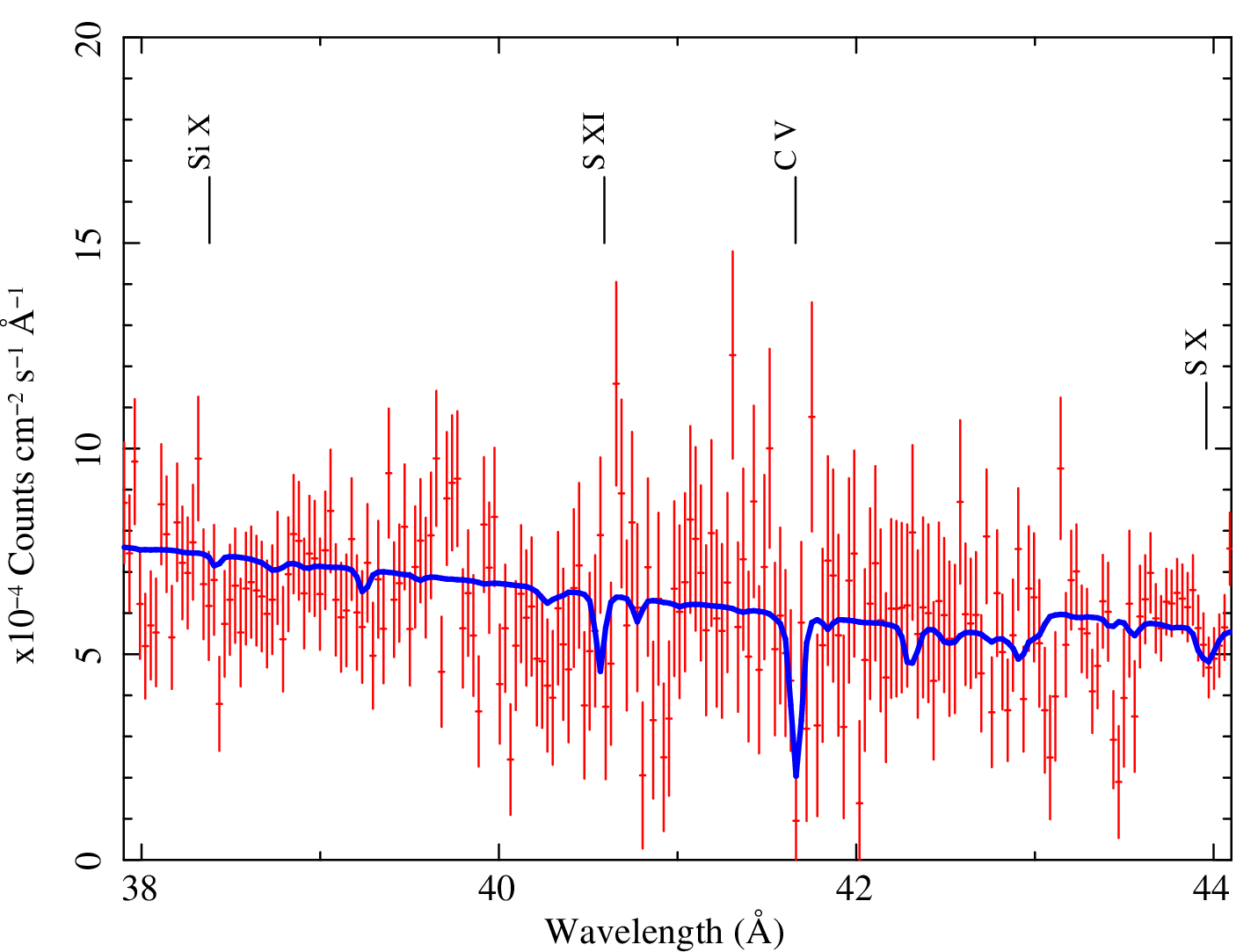}
  }
  \hbox{
    \includegraphics[height=8.5cm,width=5.25cm,angle=-90]{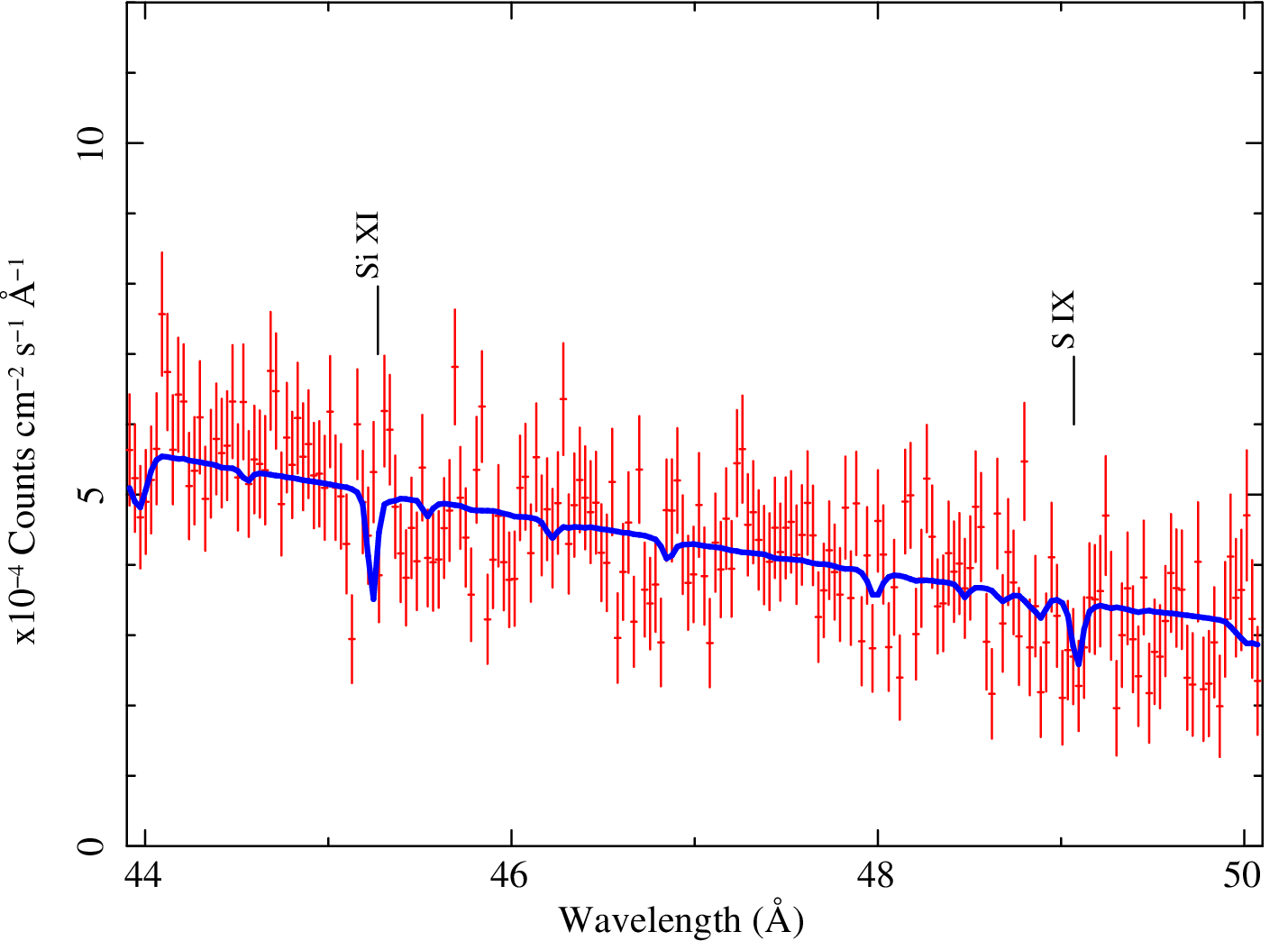}
    \includegraphics[height=8.5cm,width=5.25cm,angle=-90]{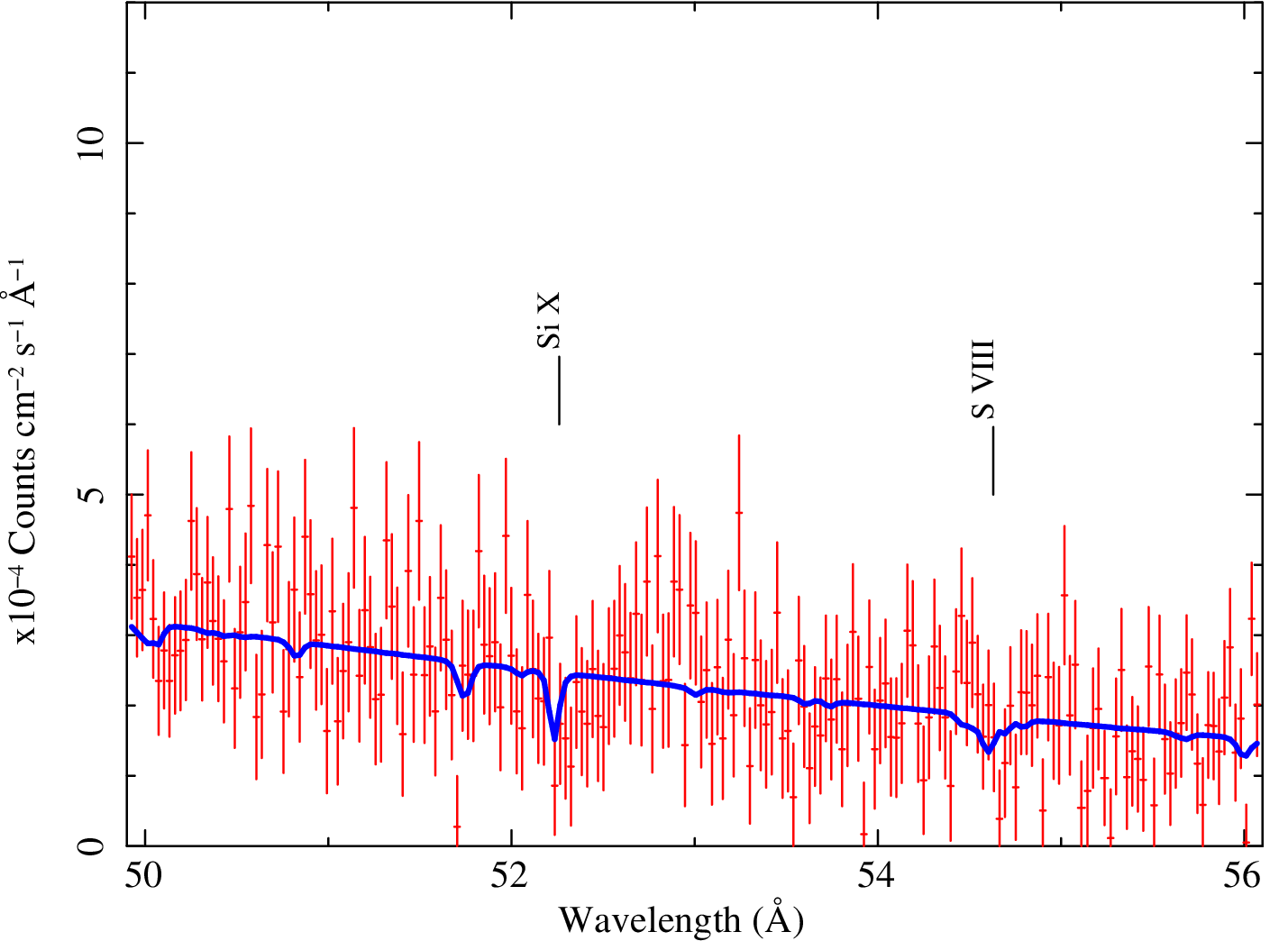}
  }
  \caption{\Chan{}-LETGS spectrum of Mrk 509. The solid line represents the best-fit model. The most prominent features are labeled.}
  \label{spec}
\end{figure*}

In summary, we observe in Mrk 509 three WA components with distinct ionization degrees (with $\log \xi = $1.06, 2.26, and 3.15, respectively). All three WA phases are somewhat shallow (with $N_{\rm H}$ a few times $10^{20}$~cm$^{-2}$), with the first component slightly redshifted with respect to the systemic velocity of Mrk 509. The second and third components are outflowing at velocities of $\sim 200$ and $\sim 460$~km~s$^{-1}$, respectively. The best-fit results are shown in Table~\ref{WApar}. The detailed LETGS spectrum along with the best-fit {\tt xabs} model is shown in Fig.~\ref{spec}.

This WA fit was performed with a spline continuum. Since the continuum was also modelled following a different approach (power-law plus modified black body) that yielded a similar goodness of fit (as explained in Sect~\ref{cont}), we repeated the WA analysis using this continuum model for comparison. In this fit, the {\it rms} velocity broadening $\sigma$ was kept fixed to those values obtained in the fit using a spline as a continuum. As can be seen in Table~\ref{WApar}, the WA parameters obtained are, as expected, fully consistent with those assuming a spline model for the continuum. 

Furthermore, we extended the analysis range towards longer wavelengths, trying to take advantange of the high effective area of LETGS up to $\sim$170~\AA. Unfortunately, the high background made only possible to carry out an analysis up to $\sim$60~\AA. The most relevant ions in this part of the spectrum correspond to \CV{}, \SiXI{}, \SiX{}, \SIX{}, and \SX{}. Since the bulk of the WA lines falls below 40~\AA, the WA parameters described with {\tt xabs} are tightly constrained. Therefore, an extended fit including the longer wavelengths provided results fully consistent with those reported in table~\ref{WApar}. A {\tt slab} fit for these ions provided upper limits for the majority of them with the exception of \SiX{}, \SIX{}, and \SX{} (see Table~\ref{slabpar}), albeit with large error bars.

\begin{table}
  \renewcommand{\arraystretch}{1.6}
  \centering
  \caption[]{Warm absorber best-fit parameters.}
  \label{WApar}
  \begin{tabular}{l c c c c c}
    \hline\hline
    Continuum  & Comp. & $\log \xi$\tablefootmark{a} & $N_{\rm H}$\tablefootmark{b} & $\sigma$\tablefootmark{c} & $v_{\rm out}$\tablefootmark{d} \\
    \hline
    Spline       &     1  & 1.06$\pm$0.13 & 1.9$\pm$0.5 & 86$^{+53}_{-19}$ & 73$^{+41}_{-59}$ \\
                 &     2  & 2.26$\pm$0.07 & 5.4$\pm$1.6 & 49$^{+62}_{-15}$ & $-196^{+87}_{-73}$ \\
                 &     3  & 3.15$\pm$0.15 & 5.9$^{+5.1}_{-2.6}$ & 75$^{+303}_{-58}$ & $-455^{+103}_{-235}$ \\
    \hline
    PL$+$mbb     &     1  & 1.06$\pm$0.14 & 1.7$\pm$0.4 & 86 $(f)$ & 73$\pm$38 \\
                 &     2  & 2.26$\pm$0.07 & 5.1$\pm$1.3 & 49 $(f)$ & $-202^{+52}_{-64}$ \\
                 &     3  & 3.35$\pm$0.27 & 8.1$^{+12.4}_{-5.6}$ & 75 $(f)$ & $-445^{+129}_{-196}$ \\
    \hline
  \end{tabular}
  \tablefoot{
  \tablefoottext{a}{Ionization parameter, in units of erg~s$^{-1}$~cm}; \\
  \tablefoottext{b}{Column density, in units of $10^{20}$~cm$^{-2}$}; \\
  \tablefoottext{c}{{\it rms} velocity broadening, in units of km~s$^{-1}$, $(f)$ denotes a fixed parameter during the fit}; \\
  \tablefoottext{d}{Outflow velocity, in units of km~s$^{-1}$.}
  }
\end{table}


\section{Discussion}
\label{discussion}

\subsection{Comparison with previous X-ray observations}
\label{RGS}

Prior to this LETGS observation, Mrk 509 has been analyzed using various high-resolution X-ray spectra. It was observed by \XMM{} in October 2000 for $\sim$30~ks. Pounds et al. (\cite{Pou01}) performed an analysis of the RGS spectrum that showed hints of intrinsic absorption, mainly \NeIX{} features possibly blended with \FeXIX{}, as well as marginal emission features in the \OVII{} complex. An additional $\sim$30~ks \XMM{} observation was made in April 2001. Smith et al. (\cite{Smi07}) combined both RGS datasets in order to increase the signal-to-noise ratio. They found a few emission lines (a broad \OVII{} complex and \CVI{} Ly$\alpha$ line, plus a narrow \OVIII{} Ly$\alpha$ line), and several absorption features that they attribute to a multi-component WA. They fitted three {\tt xabs} components to the combined RGS spectra obtaining ionization parameters fully in agreement with ours ($\log \xi = 0.89^{+0.13}_{-0.11}$, $2.14^{+0.19}_{-0.12}$, and $3.26^{+0.18}_{-0.27}$, respectively). However, there are interesting differences between the Smith et al. (\cite{Smi07}) analysis and ours. First, we find a correlation between the ionization degree and the outflow velocity, i.e the more ionized is the component, the faster it flows. Smith et al. (\cite{Smi07}) report the opposite,  with their low ionization component having a faster outflow velocity. Further, their reported column densities for two of the three components are almost one order of magnitude higher than ours. It is difficult to ascertain whether these values correspond to intrinsic variations in the properties of the WA with respect to our observation. Some probable explanations of these discrepancies are discussed in Detmers et al. (\cite{Det10}). Given the fact that the ionizing luminosity and the ionization parameter in the Smith et al. (\cite{Smi07}) analysis are comparable to ours, we speculate that their higher column densities are due to their extremely low velocity broadening, which they find close to 0, probably because of the low signal-to-noise ratio of their data. In fact, their second WA component, for which they report a $rms$ velocity of $70^{+70}_{-30}$~km~s$^{-1}$, is fully consistent with ours.

To further test this possibility, already suggested in Detmers et al. (\cite{Det10}), starting from our best-fit values using a power-law as the continuum model (for consistency with the continuum model of Smith et al. \cite{Smi07}) we have set to zero the velocity broadening of our components 1 and 3 and re-fitted the data. This renders $N_H = 2.0^{+2.7}_{-1.2}\times 10^{21}$~cm$^{-2}$ for the high-ionization component, which is in agreement with the value of Smith et al. (\cite{Smi07}) at the 1$\sigma$ confidence level. The column density of the low-ionization component, however, does not change substantially from its original value, remaining significantly different from of Smith et al. (\cite{Smi07}). Repeating this exercise with a spline model for the continuum we find a column density for the lowly ionized component which is consistent with that for the power-law model.

Mrk 509 was observed in April 2001 for $\sim$59~ks with \Chan{}-HETGS simultaneously with HST-STIS. Yaqoob et al. (\cite{Yaq03}) fitted the observed absorption lines using a single photoionization model with $\log \xi = 1.76^{+0.13}_{-0.14}$ and $N_H=2.06^{+0.39}_{-0.45}\times 10^{21}$~cm$^{-2}$. This best-fit has parameters between our components 1 and 2. However, it has a larger column density (see Table~\ref{WApar}). Interestingly, they found this component to be outflowing at $\sim$200~km~s$^{-1}$, in good agreement with the outflow velocity of our component 2. We note that this velocity was not a free parameter in their XSTAR modelling, but was introduced as a nominal velocity offset. The chosen value was based on the velocity profile analysis of the most relevant absorption lines present in the HEGTS spectrum. There are hints of a more complex kinematic structure since some high-ionization lines, such as \MgXI{} and \NeX{}, show larger blueshift values (e.g. $\sim -620$~km~s$^{-1}$ for \MgXI{}) but the uncertainties make it difficult to ascertain whether there is an actual higher velocity component. Yaqoob et al. (\cite{Yaq03}) also applied a curve of growth analysis to deduce a velocity width of 100~km~s$^{-1}$, quite similar to what we find and fully in agreement with the recent analysis of the data gathered during our 2009 multi-wavelength campaign (Kaastra et al. \cite{Kaa11b}), hereafter Paper II. The velocity widths found in X-rays are, in general, broader than those found in the UV (Kriss et al. \cite{Kriss00}, Kraemer et al. \cite{Kra03}) which may provide some hints on the nature of the X-ray/UV absorbing gas. Note, however, that the UV absorption is characterized by more components at velocities too close to be resolved in the X-rays. This may explain the widths measured in the \Chan{} spectra if there are multiple X-ray absorption systems corresponding to each of those seen in the UV (see Sect.~\ref{UV}).

Another set of three \XMM{} observations were taken in 2005 and 2006. Cappi et al. (\cite{Cap09}) used the pn spectra in these datasets and obtained evidence for a highly ionized and mildly relativistic outflow, while Detmers et al. (\cite{Det10}) used the RGS spectra to further analyze the WA in Mrk 509. They confirmed the presence of three ionization components outflowing at distinct velocities, with $\log \xi = 1.0\pm0.2$, $\log \xi = 1.95\pm0.02$, and $\log \xi = 3.20\pm0.08$, respectively, by stacking the three RGS spectra. The most remarkable difference with our results is that the outflow velocities measured by Detmers et al. (\cite{Det10}) are not correlated with the ionization parameter, finding that the intermediate $\xi$ component is roughly at the systemic velocity of the source, while the low- and high-ionization components are outflowing at $-120\pm60$ and $-290\pm40$~km~s$^{-1}$, respectively. An analysis of the individual RGS spectra yielded similar results albeit with larger error bars. Interestingly, they found a hint of changes in the WA between the 2005 and 2006 observations, taken $\sim$6~months apart. Unfortunately the statistics prevented them from constraining the location of the different WA components, except for the high-ionization one which was found to lie within 4~pc from the ionizing source.

As part of our 2009 multi-wavelength campaign, Mrk 509 was observed again by \XMM{} for a total of 600~ks, split in 10 observations of 60~ks each taken 4 days apart in October-November 2009. The detailed analysis of the stacked RGS spectrum is reported in an earlier paper in this series (Detmers et al. \cite{Det11}, hereafter Paper III). Although the sensitivity of the stacked 600~ks RGS spectrum is much higher than our LETGS observations, the results between the WA analyses of both observations are fully consistent within the error bars. The most relevant result is that this deep \XMM{} observation finds no evidence that the ionized outflows consist of a continuous distribution of ionization states, but rather they are made of discrete components. Indeed, Paper III finds 6 ionization components (two for the slow outflow, consistent with the systemic velocity of Mrk 509, and 4 for the fast outflow). If we compare our best-fit parameters reported in Table~\ref{WApar} with those of Table~5 in Paper III we clearly identify their component B1 with our component 1, whereas their component E2 corresponds with our component 3. On the other hand, our component 2 seems to be a blend of their components C2 and D2, which are unresolved in this \Chan{} dataset.

Note, however, that the outflow velocities measured in this LETGS spectrum are consistently larger than those reported for RGS in Paper III for both {\tt xabs} and {\tt slab} fits. An analysis of the wavelength scale was carried out in Paper II, where a best-fit observation/model wavelength shift of $-1.8 \pm 1.8$~m\AA~was reported combining all the indicators (see Table~11 in Paper II). In the same table, the difference in the wavelength scale between the RGS observation and the LETGS observation is quoted as $-7.3 \pm 3.8$~m\AA. Hence, there is a difference of $\sim$5.5~m\AA~ between the adopted wavelength in RGS and that of LETGS, the latter being larger (redshifted) with respect to the former. At a typical wavelength of 20~\AA, 5.5~m\AA~corresponds to a velocity offset of 83~km~s$^{-1}$. Therefore this offset should in principle be substracted to the outflow velocities quoted in Tables~\ref{slabpar}~and \ref{WApar} if one wants to compare with the equivalent models reported in Paper III. However, there is, within the systematic uncertainties, {\it no discrepancy} between RGS and LETGS wavelength scales.

Interestingly, the average flux state of Mrk 509 during the \XMM{} observation was $\sim$60\% higher than during the \Chan{} observation in the soft X-ray waveband ($20-35$~\AA), and $\sim$30\% higher in the hard ($7-10$~\AA) band. Since the maximal variation happens at low energies we expect very little effect on most of the ionization states observed. The variations at harder energies may have an effect but the WA parameters remain similar in both observations. It is then difficult to state whether this is due to an intrinsic lack of variation of the gas (i.e. the gas did not have time to recombine after the flux decrease) or that there is an actual change in the parameters but it is statistically undetectable. Indeed, errors for some of the key parameters such as the column densities, which would allow us to put some constraints on the location of the WA, can be as high as $\sim$50\%.

\subsection{Structure and location of the warm absorber}
\label{structure}

In this section we discuss the nature and location of the ionized absorbers of Mrk 509. The low-ionization phase (component 1 in Table~\ref{WApar}) appears to be slightly redshifted and is consistent with being at the systemic velocity at the $2\sigma$ level, while components 2 and 3 are outflowing at several hundreds of km~s$^{-1}$. This suggests the possibility that component 1 is part of a different structure than components 2 and 3. To test this we plot the pressure ionization parameter $\Xi$ as a function of the electron temperature $T$. Components sharing the same $\Xi$ value in this thermal stability curve would likely be part of the same long-lived structure. The pressure ionization parameter is defined as $\Xi = L/4\pi r^2 cp = \xi/6\pi ckT$. It was calculated using {\tt CLOUDY}, which provided a grid of ionization parameters $\xi$ and the corresponding electron temperatures for a thin layer of gas illuminated by an ionizing continuum. The resulting curve divides the $T-\Xi$ plane in two regions: below the curve heating dominates cooling, whereas above the curve cooling dominates heating (e.g. Reynolds \& Fabian \cite{RF95}, Chakravorty at al. \cite{Cha08}). Therefore, the parts of the curve where $dT/d\Xi<0$ (where the curve turns back) are unstable against isobaric thermal perturbations (vertical displacements from the curve), while those with positive derivative $dT/d\Xi$ are stable.

\begin{figure}
  \centering
  \includegraphics[width=7cm,angle=-90]{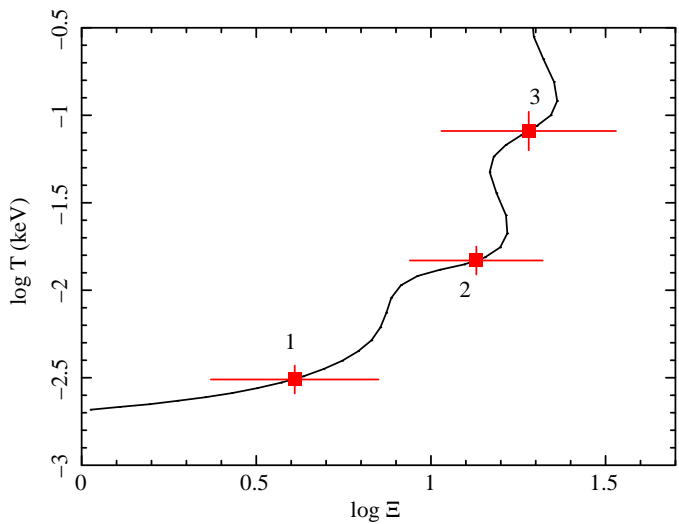}
  \caption{Pressure ionization parameter as a function of electron temperature. The filled squares represent the warm absorber components in Mrk 509.}
  \label{xitemp}
\end{figure}

If we overplot the WA components on top of this stability curve (see Fig.~\ref{xitemp}) we can see that the three of them fall in stable areas, which suggests that the absorbing gas belongs to a discrete scenario rather than being part of a continuous distribution of ionized material. Furthermore, the value of $\Xi$ for components 2 and 3 are consistent within the error bars. This would mean that both components nominally share the same pressure ionization parameter, i.e. they are in pressure equilibrium. On the other hand, component 1 has a significantly different $\Xi$ value which, in addition to the fact that its features are not blueshifted, strongly suggests that it may not coexist as a part of the same outflow structure as components 2 and 3. This will be further discussed in Sect.~\ref{UV}.

The location of the WA is hard to estimate since the product $nR^2$ is degenerate. The best way to put constraints on the distance $R$ is therefore by using variability to determine the densities of the WA (e.g. Longinotti et al. \cite{Lon10}, Detmers et al. \cite{Det08}). As mentioned in Sect.~\ref{RGS}, there was a variation in flux of $\sim$30\% between the \XMM{} 600~ks observation and the \Chan{} observation that would help to constrain the distance if changes in the WA were detectable. For that purpose we compared the measured column densities of the different ionization states of oxygen, as these ions were the most accurately measured in both datasets. At a first glance the ionic column densities measured in both cases are consistent within the error bars (see Table~\ref{slabpar}, and Table~4~in Paper III). We statistically tested this by applying a two-dimensional Kolmogorov-Smirnov test on both ionic column density distributions of oxygen in the same manner as in Ebrero et al. (\cite{Ebr10}). The test was performed for components 1 and 2 in model 2 of Table~\ref{slabpar} and its equivalent in Paper III. The significance of the variation in component 1 between both datasets was at the $\sim$90\% confidence level, while for component 2 it was well below the 1$\sigma$ level. Therefore, there are no grounds to estimate the distance of the WA absorber based on variability.

Alternatively, assuming that the outflows are likely to be filamentary or clumpy, a volume filling factor defined in the same manner as in Blustin et al. (\cite{Blu05}) can be characterized so that $C_{\rm V} < 1$. Therefore, if we integrate the density along our line of sight, we get $N_{\rm H}=C_{\rm V}nR$, where $n$ is the density at the base of the absorber. Substituting $n$ in the definition of the ionization parameter $\xi$ we obtain

\begin{equation}
\label{R}
R\leq \frac{LC_{\rm V}}{\xi N_{\rm H}}.
\end{equation}

\noindent We can estimate from our spectral fits the values of $\xi$, $N_{\rm H}$, and the ionizing $1-1000$~Ryd luminosity $L$, whereas the volume filling factor can be estimated following dynamical arguments, assuming that the momentum of the outflow must be close to the momentum of the absorbed radiation plus the momentum of the scattered radiation (Blustin et al. \cite{Blu05}):

\begin{equation}
\label{cv}
C_{\rm V}\sim \frac{(\dot{P}_{\rm abs}+\dot{P}_{\rm sc})\xi}{1.23m_{\rm p}Lv^2\Omega},
\end{equation}

\noindent where $\dot{P}_{\rm abs}$ and $\dot{P}_{\rm sc}$ are the momenta of the absorbed and scattered radiations, respectively, $m_p$ is the mass of the proton, and $\Omega$ is the solid angle subtended by the outflow. We can estimate both momenta using formulas 20 and 21 in Blustin et al. (\cite{Blu05}), and we take $\Omega = \pi$~sr, inferred from the type-1/type-2 ratio in nearby Seyfert galaxies of Maiolino \& Rieke (\cite{MR95}). Using an ionizing luminosity of $L=3.2\times 10^{45}$~erg~s$^{-1}$ derived from the average SED of Mrk 509, and WA best fit values reported in Table~\ref{WApar}, we obtain volume filling factors $C_{\rm V}$ of $\sim$0.01, $\sim$0.03 and $\sim$0.04 for components 1, 2 and 3 in the WA, respectively. Using now Eq.~\ref{R} we find that the outflowing components 2 and 3 are located within 300 and 50~pc, respectively, from the central ionizing source. Using the relation of Krolik \& Kriss (\cite{KK01}) the inner edge of the torus is estimated to be at $\sim$5.5~pc from the ionizing source. With the rather mild limits derived here is difficult to constrain the location of the WA, but it is consistent with the possibility that the outflows are generated in the putative torus, although an inner location cannot be ruled out. For component 1, which is probably not part of the outflow, we derive a much larger upper limit of $R \lesssim 6$~kpc, giving support to the idea that this component probably belongs to the ISM of the host galaxy.

In terms of cosmic feedback, we can estimate the mass outflow rate for components 2 and 3 using Eq.~19 in Blustin et al. (\cite{Blu05}):

\begin{equation}
\label{eq19}
\dot{M}v \sim \dot{P}_{\rm abs}+\dot{P}_{\rm sc}.
\end{equation}

\noindent In this way we obtain $\dot{M}_2 \sim 1$~M$_{\odot}$~yr$^{-1}$ and $\dot{M}_3 \sim 0.4$~M$_{\odot}$~yr$^{-1}$. The total kinetic energy carried by the outflow per unit time (the sum of the kinetic luminosity of the blueshifted components), which can be defined as $L_{\rm KE}=1/2 \dot{M}_{\rm out} v^2$, is of the order of $4 \times 10^{40}$~erg s$^{-1}$ which represents a tiny fraction of the bolometric luminosity of Mrk 509, approximately $L_{\rm KE}/L_{\rm bol} \sim 0.001\%$. It is then very unlikely that these outflows can affect significantly the local environment of the host galaxy. Indeed, current AGN feedback models that reproduce the $M - \sigma$ relation predicts that $\sim$5\% of the AGN radiative output is required to be fed back as both kinetic and internal motion energy in the form of outflows in order to significantly affect the immediate surroundings of the central engine (e.g. Silk \& Rees \cite{SR98}, Wyithe \& Loeb {\cite{WL03}}, Di Matteo, Springel \& Hernquist \cite{DSH05}). The velocity widths that we measure are of the order of 100~km~s$^{-1}$, approximately the same order of magnitude as the outflow velocity. The internal motion energy of the gas is hence similar to its bulk kinetic energy but, nevertheless, it is still far too low to generate a significant impact even if we consider a less conservative energetic requirements to disrupt the local ISM of Mrk 509 (Hopkins \& Elvis \cite{HE10}). Unless a higher velocity and mass outflows are present (albeit undetected in the LETGS, but see Sect. ~\ref{UFO}), the shallow and relatively slow outflows detected in Mrk 509 are therefore not powerful enough to play an important role in cosmic feedback.

\begin{figure}
  \centering
  \includegraphics[width=6.5cm,angle=-90]{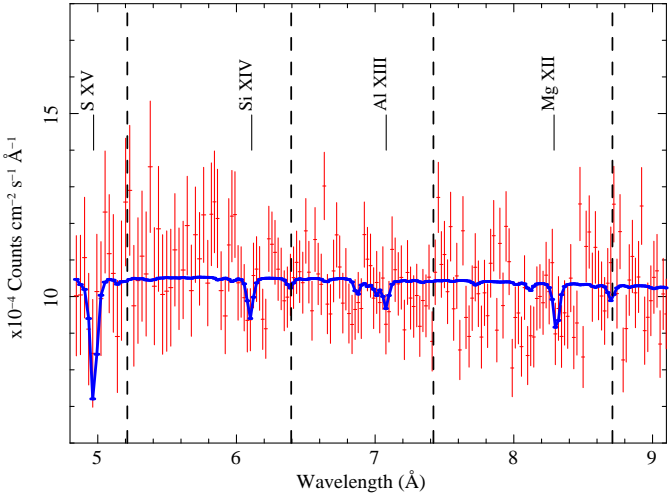}
  \caption{\Chan{}-LETGS spectrum of Mrk 509 in the $\sim 5-9$~\AA region. The solid line represents the best-fit model. The most prominent features are labeled. Note the shift of the troughs with respect to their wavelengths at the rest frame of the source (vertical dashed lines).}
  \label{ufospec}
\end{figure}

\subsection{Is there an ultra-fast outflow?}
\label{UFO}

The region of the spectrum at shorter wavelengths, between $\sim$5 and 9~\AA, shows a noisy continuum although some absorption troughs could be present. The most prominent ones were those at $\sim$4.95~\AA~and $\sim$6.1~\AA. A tentative identification of these lines suggested that they could be \SXV{} and \SiXIV{}, respectively. If this was the case that would mean that the gas responsible for these features is outflowing at $\sim 14,000$~km~s$^{-1}$. We tested this possibility by fitting a {\tt slab} model in which the outflow velocity and the ionic column densities of \SXV{}, \SiXIV{}, \AlXIII{}, and \MgXII{} (the most prominent lines expected in this range of the spectrum) were the free parameters, while the widths of the lines were kept frozen to 100~km~s$^{-1}$. The addition of this component improved the overall fit of the spectrum only at the 90\% confidence level and yielded an outflow velocity of $-14,500^{+560}_{-410}$~km~s$^{-1}$ (see Table~\ref{ufopar}). The spectrum along with the best-fit model is shown in Fig.~\ref{ufospec}.

Ponti et al. (\cite{Pon09}) also reported the presence of a possible ultra-fast outflow in Mrk 509 based on the analysis of highly ionized Fe absorption lines present in the \XMM{} EPIC spectra of 2005 and 2006. They obtained an outflow velocity of $\sim -14,500\pm 3600$~km~s$^{-1}$, which would be in agreement with our marginal detection. They predict, however, that the outflowing gas should be highly ionized ($\log \xi = 5.15$), whereas the concentration of the ions in our detection peaks at values of $\log \xi$ in the range 2.7-3.1 and would be hardly detectable at much higher ionization states. The \Chan{}-HETGS spectrum of Mrk 509 reported in Yaqoob et al. (\cite{Yaq03}) does not show significant absorption troughs at these wavelengths and therefore we are unable to confirm the presence of such an ultra-fast outflow.

\begin{table}
  \centering
  \caption[]{Best-fit parameters of the tentative ultra-fast outflow.}
  \label{ufopar}
  \begin{tabular}{l c c}
    \hline\hline
    $v_{\rm out}$\tablefootmark{a} &  $-14,500^{+560}_{-410}$ &   \\
    \hline
    Ion  &  $\log N_{\rm ion}$\tablefootmark{b} & $\log \xi$\tablefootmark{c} \\
      \hline
      \MgXII{}  & 16.57$\pm$0.34 & 2.75 \\
      \AlXIII{} & $<$16.7  &  2.90 \\
      \SiXIV{}  & 16.71$\pm$0.39 & 3.05 \\
      \SXV{}  & 17.65$\pm$1.20 & 2.80 \\
      \hline
  \end{tabular}
  \tablefoot{
    \tablefoottext{a}{Outflow velocity, in units of km~s$^{-1}$;} \\
    \tablefoottext{b}{Ionic column density, in units of cm$^{-2}$;} \\
    \tablefoottext{c}{Ionization parameter at which the ion fraction peaks for each ion, in units of erg~s$^{-1}$~cm.}
    }
\end{table}

\subsection{Connection with the UV}
\label{UV}

\begin{table*}
  \centering
  \caption[]{Properties of the \COS{} intrinsic features.}
  \label{UVprop}
  \begin{tabular}{l c c c c c}
    \hline\hline
    UV Component & $v_{\rm out}$\tablefootmark{a} & $N_{\rm CIV}$\tablefootmark{b} & $N_{\rm NV}$\tablefootmark{b} & $N_{\rm OVI}$\tablefootmark{b,c} & LETGS\tablefootmark{d} \\
    \hline
1       &  $  -408 \pm   5$ & $  31.2\pm  1.5$ & $ 107.0\pm  9.5$ & $ 215.0\pm 47.2$   & 3 \\
1b      &  $  -361 \pm  13$ & $  16.3\pm  6.1$ & $  17.5\pm  1.6$ & $ 154.9\pm 39.0$   & 3? \\
2       &  $  -321 \pm   5$ & $ 136.3\pm 41.7$ & $ 149.0\pm 10.7$ & $ 566.2\pm 118.0$  & 3? \\
2b      &  $  -291 \pm   6$ & $ 128.9\pm 41.4$ & $ 130.6\pm 11.3$ & $1248.1\pm 395.5$  & 2? \\
3       &  $  -244 \pm   5$ & $  47.7\pm  7.2$ & $  88.6\pm  8.1$ & $ 675.2\pm 157.1$  & 2 \\
4a      &  $   -59 \pm   5$ & $  66.3\pm  2.1$ & $  93.6\pm  6.8$ & $ 804.5\pm 387.2$  & \\
4       &  $   -19 \pm   5$ & $ 250.0\pm 11.0$ & $ 323.6\pm 12.2$ & $8797.0\pm 4120.3$ & \\
5       &  $    37 \pm   5$ & $  36.2\pm 16.9$ & $  38.3\pm 11.1$ & $ 683.5\pm 429.3$  & 1 \\
6a      &  $    79 \pm  12$ & $   5.6\pm  2.3$ & $  20.6\pm  6.4$ & $ 518.8\pm 398.5$  & 1 \\
6       &  $   121 \pm   5$ & $  12.3\pm  6.4$ & $  56.1\pm  4.0$ & $3436.1\pm 6985.0$ & 1? \\
6b      &  $   147 \pm   8$ & $  17.6\pm  5.3$ & $   5.5\pm  1.8$ & $ 104.5\pm 234.6$  &  \\
7a      &  $   184 \pm   6$ & $   3.6\pm  1.1$ & $  21.0\pm  2.9$ & $ 660.2\pm 176.7$  &  \\
7       &  $   222 \pm   6$ & $   6.4\pm  0.8$ & $  23.5\pm  2.7$ & $ 667.4\pm 1105.3$ &  \\
    \hline
  \end{tabular}
  \tablefoot{
    \tablefoottext{a}{Outflow velocity, in units of km~s$^{-1}$;} \\
    \tablefoottext{b}{Ionic column density, in units of $10^{12}$~cm$^{-2}$;} \\
    \tablefoottext{c}{Based on {\it FUSE} observations;} \\
    \tablefoottext{d}{LETGS WA component counterpart.} \\
  }
\end{table*}

Prior to this campaign, Mrk 509 was observed in the UV domain by FUSE (Kriss et al. \cite{Kriss00}) in 1999, and by HST-STIS (Kraemer et al. \cite{Kra03}) in 2001, the latter simultaneously with \Chan{}-HETGS (Yaqoob et al. \cite{Yaq03}). Their spectra showed the presence of seven and eight different kinematic components, respectively, with approximately half of them blueshifted and the other half redshifted with respect to the systemic velocity of the source.

As a part of our multiwavelength campaign, Mrk 509 was observed with \COS{} simultaneously with \Chan{}-LETGS. The data reduction processes and discussion of the results of this observation are presented in a companion paper (Kriss et al. \cite{Kriss11}).

Similarly to previous observations, \COS{} observations find that the UV absorption lines in Mrk 509 present 13 distinct kinematic components, ranging from $-$408 to $+$222~km~s$^{-1}$ (the minus sign denotes an outflow). For consistency, we list the UV components in this paper using the same numbering as in Kriss et al. (\cite{Kriss11}), component 1 being the most blueshifted and component 7 the most redshifted. Note that Kriss et al. (\cite{Kriss11}) numbered the UV components using kinematic criteria, while for the X-ray components we have adopted a numbering criterion based on the ionisation state of the gas. In Table~\ref{UVprop} we list the main properties of the intrinsic absorption features in the \COS{} spectra, such as the velocity shift with respect to the systemic velocity and the ionic column densities measured for the \CIV{}, \NV{} and \OVI{} ions, as well as the possible LETGS X-ray counterparts. Note that the \COS{} observation was taken in the $1160-1750$~\AA~spectral range. Since the \OVI{} feature falls outside of this range, the data relevant to this ion are those of the previous {\it FUSE} observation of Mrk 509 (Kriss et al. \cite{Kriss00}).

\subsubsection{Kinematics}
\label{UVkinematics}

Attending to their kinematic properties, one can see that the gas responsible for the UV absorption and the X-ray absorbing gas share some of the components, suggesting that they might be co-located. For instance, the UV components 1 and 1b, outflowing at $-$408 and $-$361~km~s$^{-1}$, respectively, may correspond with the X-ray component 3. We note, however, that the resolution in the X-ray domain is not comparable with that of the UV, so a one-to-one component identification is not always possible. Indeed, some of the X-ray kinematic components may be a blend of velocities and ionization states (see Paper III). Within the error bars, X-ray component 3 may also be partly identified with UV component 2. On the other hand, there is a clear correspondence between UV component 3, outflowing at $-$244~km~s$^{-1}$, and possibly UV component 2b at $-$291~km~s$^{-1}$, with the intermediate X-ray component 2. The UV components 5, 6 and 7 are inflowing, their radial velocity redshifted with respect to the source rest frame, similarly as the X-ray component 1. Both UV components 5 and 6a, and X-ray component 1 are flowing close to the systemic velocity of Mrk 509, which may indicate that they move mainly transverse to our line of sight.

Since the inflow velocities of the UV components 6, 6b, 7a, and 7 are of the order of hundreds of km~s$^{-1}$, and the redshifted X-ray component is not in pressure equilibrium with the others, strongly suggests that these components are not part of the outflow and may have an extra-nuclear origin. In fact, Kriss et al. (\cite{Kriss11}) speculate with the possibility that we are detecting high-velocity clouds in the interstellar medium of Mrk 509 host galaxy, similar to the Complex C in our own Galaxy (Thom et al. \cite{Thom08}).

\subsubsection{Column densities}
\label{UVdensities}

The total ionic column densities of \CIV{} and \NV{} measured in the \COS{} spectra summed over all the kinematic components are of the order of $\log N_{\rm ion}~({\rm cm}^{-2}) \sim 15$, so they are not likely to produce substantial X-ray absorption. Indeed, our X-ray analysis yields an upper limit for $\log N_{\rm NV} < 15.4$ (see Table~\ref{slabpar}) which seems consistent with the ionic column density measured by \COS{} but, however, cannot be used to probe if the UV/X-ray absorbers have a common origin. On the other hand, the \OVI{} measurement based on {\it FUSE} observations in 1999 provided higher column density which therefore could be also detectable in X-rays.

Making direct comparisons between the UV observations of \OVI{} and the X-ray, even among the individual \OVI{} components seen with {\it FUSE} is difficult because of the high level of saturation in most of these features, as pointed out by Kraemer et al. (\cite{Kra03}). The quoted column densities from Kriss et al. (\cite{Kriss00}) for the strongest components should be treated as lower limits. The deepest \OVI{} absorption is seen in UV component 4, which is at rest at the systemic velocity of the source (see Table~\ref{UVprop}). No X-ray component is found at this velocity in our LETGS spectrum. Other strong (and saturated) \OVI{} absorption components are seen in UV components 3 and 6, which likely correspond to X-ray components 2 and 1, respectively. The latter is redshifted and is probably not part of the outflow, but the former shares the same kinematic properties as X-ray component 2 so they are likely co-located. We note that the {\it FUSE} and LETGS observations were taken almost 10 years apart, and therefore variations in the \OVI{} column densities, although unlikely, are possible (Kriss et al. \cite{Kriss11}).

This poses the problem of disentangling whether the gas responsible for the UV absorption and the X-ray absorption are the same. Kraemer et al. (\cite{Kra03}) claimed that none of the UV absorbers detected in the {\it HST-STIS} spectrum of Mrk 509 could be identified with the HETGS X-ray absorbing gas described in Yaqoob et al. (\cite{Yaq03}), as the sum of all the ionic column densities of \OVII{} spread over the different UV kinematic components only accounted for the 10\% of the column density for the same ion measured in the HETGS spectrum. The total value reported in Kraemer et al. (\cite{Kra03}) is $\log N_{\rm OVII} = 16.51$, which is indeed only 25\% of the $N_{\rm OVII}$ column density we measure in our LETGS spectrum assuming that this feature originates in a single kinematic component (model 1 in Table~\ref{slabpar}) as in Yaqoob et al. (\cite{Yaq03}). When we consider a multi-velocity wind (model 2 in Table~\ref{slabpar}) then the \OVII{} column densities measured in either kinematic components are in agreement with the {\it HST-STIS} result within the error bars. Likewise, if we follow the same argument considering the total \OVI{} column density measured by {\it FUSE}, which sums up to $\log N_{\rm OVI} = 16.1$ and should be considered as a lower limit, we find that this result is fully in agreement with ours when we assume a single kinematic component. More interestingly, the \OVI{} column density in the UV component 3 ($\log N_{\rm OVI} = 15.4$, also a lower limit) is consistent within the error bars with the measured column density in LETGS for the corresponding X-ray kinematic component ($\log N_{\rm OVI} = 15.7\pm0.4$). 

We see that the absorbing gas scenario in Mrk 509 is quite complex. The results obtained in this campaign show that the UV and X-ray absorbers have consistent column densities. Kriss et al. (\cite{Kriss11}) shows that the UV absorbers, although characterized in discrete kinematic components, are likely to be smoothly distributed over a range in velocities, in particular for highly ionized species. If the UV- and X-ray-absorbing gases are indeed co-located, at least for some of the detected components, this could be consistent with a clumpy scenario in which high-density low-ionization UV-absorbing clouds are embedded in a low-density high-ionization X-ray gas, as proposed in Kriss et al. (\cite{Kriss00}) and (\cite{Kriss11}). We defer a more exhaustive analysis on the relation between the X-ray and UV absorbing gas including all the observations available from this campaign (\Chan{}-LETGS, \XMM{} RGS, and \COS{}) to a forthcoming paper (Ebrero et al., in preparation).


\section{Conclusions}
\label{conclusions}

We have presented here the results from the 180~ks \Chan{}-LETGS spectrum of Mrk 509 as part of an extensive multi-wavelength campaign. We found that the ionized absorber intrinsic to the source presents three distinct degrees of ionization ($\log \xi = $1.06, 2.26, and 3.15, respectively) and is somewhat shallow ($N_{\rm H} = $ 1.9, 5.4, and 5.9 times $10^{20}$~cm$^{-2}$, respectively). The low-ionization component is slightly redshifted ($\Delta v = 73$~km~s$^{-1}$), whereas the others are blueshifted with respect to the systemic velocity of the source ($\Delta v = -196$ and $-$455~km~s$^{-1}$, respectively, see Table~\ref{WApar}), and therefore they are outflowing. Their outflow velocities are correlated with the ionization parameter so that the higher the ionization parameter the faster is the outflow. The redshifted component is not in pressure equilibrium with the outflowing components, thus indicating that it may not form part of the same structure. We put mild constraints on the location of the outflowing gas, which would be consistent with a torus wind origin, although an inner origin cannot be ruled out. The energetics of the outflows indicate that they have a negligible effect in the interstellar medium of the host galaxy.

Mrk 509 was also observed in the UV domain with \COS{} simultaneously with \Chan{}, showing 13 kinematic components in the intrinsic UV absorption features. We compared the X-ray and UV absorbers in order to ascertain whether they have a common origin. At least three of the UV components could correspond with the X-ray components detected in the LETGS spectrum based only on the kinematics of the gas, which points to a possible co-location of both absorbing gases. The column density of \OVI{} measured in one of the X-ray components is in agreement with that of a previous UV observation with {\it FUSE} for their corresponding kinematic component. Similarly, the upper limits on the column densities of \CIV{} and \NV{} found in X-rays are also consistent with the values obtained by \COS{}. Therefore, the similar kinematics and column densities indicate that the gases responsible for the X-ray and UV absorption share a common origin.

\begin{acknowledgements}
The Space Research Organisation of The Netherlands is supported financially by NWO, the Netherlands Organisation for Scientific Research. GAK and NA gratefully acknowledge support from NASA/\XMM{}~Guest Investigator grant NNX09AR01G. Support for HST program number 12022 was provided by NASA through grants from the Space Telescope Science Institute, which is operated by the Association of Universities for Research in Astronomy, Inc., under NASA contract NAS5-26555. KCS acknowledges the support of Comit\'e Mixto ESO-Gobierno de Chile. SB acknowledges financial support from contract ASI-INAF n. I/088/06/0. GP acknowledges support via an EU Marie Curie Intra-European Fellowship under contract no. FP7-PEOPLE-2009-IEF-254279. POP acknowledges financial support from CNES and French GDR PCHE. We thank the anonymous referee for his/her comments and suggestions.
\end{acknowledgements}

\end{document}